\documentclass[twocolumn,superscriptaddress,amsmath,amssymb,pra,longbibliography]{revtex4-1}

\usepackage{amssymb,amsmath}
\usepackage{cmap}
\usepackage{graphicx}
\usepackage{bm}
\usepackage{color}
\usepackage{blindtext}
\usepackage{hyperref}
\usepackage{listings}
\usepackage{booktabs}
\usepackage{multirow}
\usepackage{amsmath}
\usepackage{mathrsfs}
\usepackage{braket}

\usepackage{dsfont}
\usepackage{tabularx}

\newcommand{\red}[1]{{\color{black}{#1}}}

\begin{document}

\title{Polarization singularities and M\"{o}bius strips in sound and water-surface waves}

\newcommand{\affilB}{Universit\'{e} de Bordeaux, CNRS, LOMA, UMR 5798, Talence, France}
\newcommand{\affilRIKEN}{Theoretical Quantum Physics Laboratory, RIKEN Cluster for Pioneering Research, Wako-shi, Saitama 351-0198, Japan}
\newcommand{\affilM}{CNRS, Centrale Marseille, Institut Fresnel, Aix Marseille University, UMR 7249, 13397 Marseille CEDEX 20, France}

\author{Konstantin Y. Bliokh}
\affiliation{\affilRIKEN}

\author{Miguel A. Alonso}
\affiliation{\affilM}
\affiliation{The Institute of Optics, University of Rochester, Rochester, NY 14627, USA}

\author{Danica Sugic}
\affiliation{\affilRIKEN}

\author{Mathias Perrin}
\affiliation{\affilB}

\author{Franco Nori}
\affiliation{\affilRIKEN}
\affiliation{RIKEN Center for Quantum Computing, Wako-shi, Saitama 351-0198, Japan}
\affiliation{Physics Department, University of Michigan, Ann Arbor, Michigan 48109-1040, USA}

\author{Etienne Brasselet}
\affiliation{\affilB}



\begin{abstract}
We show that polarization singularities, generic for any complex vector field but so far mostly studied for electromagnetic fields, appear naturally in inhomogeneous yet monochromatic sound and water-surface (e.g., gravity or capillary) wave fields in fluids or gases. The vector properties of these waves are described by the velocity or displacement fields characterizing the local oscillatory motion of the medium particles. We consider a number of examples revealing C-points of purely circular polarization and polarization M\"{o}bius strips (formed by major axes of polarization ellipses) around the C-points in sound and gravity wave fields. Our results (i) offer a new readily accessible platform for studies of polarization singularities and topological features of complex vector wavefields and (ii) can play an important role in characterizing vector (e.g., dipole) wave-matter interactions in acoustics and fluid mechanics.
\end{abstract}


\maketitle

\section{Introduction}

Polarization and spin are inherent properties of vector waves. These are typically associated with classical electromagnetic/optical fields or quantum particles with spin \cite{Azzam_book,Andrews_book,BLP}. 
Recently, it was noticed that sound waves in fluids or gases \cite{Shi2019,Bliokh2019,Bliokh2019_II,Toftul2019,Rondon2019,Long2020, Burns2020} as well as water-surface (e.g., gravity) waves \cite{Bliokh2020,Sugic2020} also possess inherent vector properties, and the notions of polarization and spin are naturally involved there \red{(see also earlier Refs.~\cite{Jones1973,Longuet-Higgins1980})}. These properties are related to the wave-induced motion of the medium's particles. Such motion can be characterized by the vector velocity field ${\bm {\mathcal V}}({\bf r},t)$ or the corresponding displacement field ${\bm {\mathcal R}}({\bf r},t)$, ${\bm {\mathcal V}} = \partial_t {\bm {\mathcal R}}$, in a way entirely analogous to, e.g., the electric field ${\bm {\mathcal E}}({\bf r},t)$ or the corresponding vector-potential ${\bm {\mathcal A}}({\bf r},t)$, ${\bm {\mathcal E}} = - \partial_t {\bm {\mathcal A}}$, in an electromagnetic wave.    

The main difference between electromagnetic and sound-wave polarizations is that the former are {\it transverse} (the fields ${\bm {\mathcal E}}$ and ${\bm {\mathcal A}}$ are orthogonal to the wavevector ${\bf k}$ for a plane wave), while the latter are {\it longitudinal} (the fields ${\bm {\mathcal V}}$ and ${\bm {\mathcal R}}$ are parallel to the wavevector for a plane wave). 
In the case of gravity or capillary waves, which appear on surfaces of classical fuids or gases \cite{LLfluid}, a plane wave has \red{a {\it mixed} nature. Namely, the fields ${\bm {\mathcal V}}$ and ${\bm {\mathcal R}}$ have longitudinal components along the wavevector lying in the unperturbed water-surface plane, as well as vertical components normal to the surface and the wavevector.} Akin to other surface or evanescent waves \cite{Shi2019,Bliokh2019,Bliokh2015PR,Aiello2015}, these two components are mutualy $\pi/2$ phase-shifted, so that gravity plane waves are {\it elliptically}-polarized in the propagation plane \cite{surfacewave}.

However, when one considers structured (inhomogeneous) wave fields, consisting of many plane waves, these differences between transverse, longitudinal, and mixed plane-wave polarizations are largely eliminated. Indeed, at a given point ${\bf r}$, a vector monochromatic field, whether electromagnetic, acoustic, or water-surface, traces an ellipse which can have arbitrary orientation in 3D. Considering the spatial distribution of such ellipses across the ${\bf r}$-space, one deals with inhomogeneous polarization textures.
Important generic and topologically-robust characteristics of inhomogeneous wave fields are {\it singularities}: phase singularities in scalar fields and {\it polarization singularities} in vector polarization fields \cite{Dennis2009}.

\red{In physics of fluids, the emergence of various singularities is a longstanding problem attracting continuous interest \cite{eggers_PRFluids_2018, moffatt_PRFluids_2019}. In particular, the topological nature of singularities allows one to use these for the characterization of complex flows (e.g., vortices in turbulence). Furthermore, singularities can be closely related to the formation of robust topologically nontrivial objects, such as knots \cite{kleckner_natphys_2013,zhang_natcommun_2020} or M\"{o}bius strips \cite{chen_apl_2004,goldstein_pnas_2010}.

Therefore, it is not surprising that} both phase singularities and 2D polarization singularities \red{in wave fields} were first observed in the scalar and 2D-current representations of tidal ocean waves \cite{Whewell,Hansen,Berry2001,Nye1988}. 
However, a systematic treatment of structured wave fields has only been developed within the framework of {\it singular optics} \cite{Nye1987,Soskin2001,BerryDennis2001,Dennis2009}. According to this approach, generic singularities of 2D (paraxial) and 3D (nonparaxial) polarization fields are {\it C-points} \red{or C-lines} of purely circular polarizations as well as {\it L-lines} \red{or L-surfaces} of purely linear polarizations \cite{Nye1983, Nye1987, Hajnal1987, BerryDennis2001, Dennis2009, BAD2019}, and {\it polarization M\"{o}bius strips} \cite{Freund2010,Freund2010II,Dennis2011,Bauer2015,Galvez2017,Garcia2017,Kreismann2017, Bauer2019, Tekce2019, BAD2019} which are formed (solely in 3D fields) by major axes of polarization ellipses around C-points/lines. These objects are very robust because of their topological nature; they also have important implications in the geometric-phase and angular-momentum properties of the field \cite{BAD2019}. 

\begin{figure*}[t]
\includegraphics[width=0.85\linewidth]{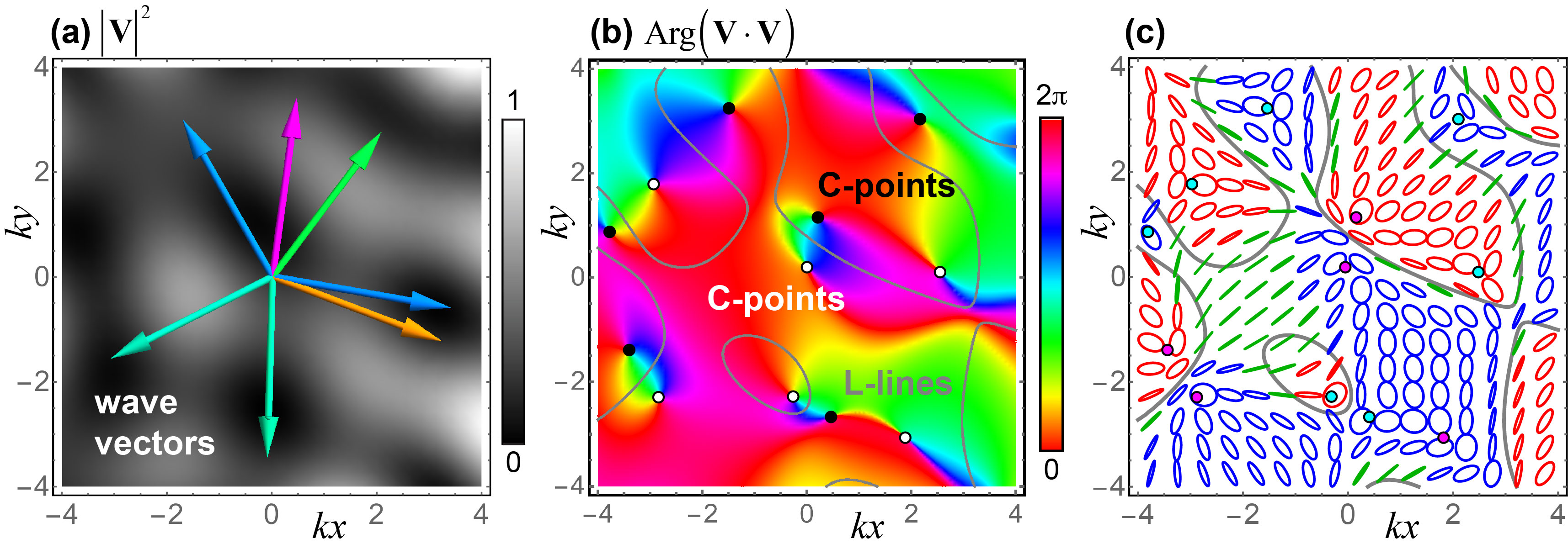}
\caption{Random 2D acoustic field obtained by the interference of $N=7$ plane waves with the same frequency and amplitude but with random directions within the $(x,y)$ plane and random phases. (a) Wavevectors of the interfering waves (with color-coded phases) and distribution of the intensity of the velocity field $|{\bf V}|^2$ (greyscale). (b) Color-coded distribution of the phase of the quadratic field ${\bf V}\cdot {\bf V}$. The phase singularities (vortices) of this field correspond to C-points of purely circular polarization in the field ${\bf V}$. L-lines correspond to purely linear polarizations. (c) Distribution of the normalized polarization of ${\bf V}$. Red, blue, and green colors correspond to right-handed, left-handed, and near-linear polarizations, respectively. The C-points in a 2D polarization field can be labeled by two independent topological numbers \cite{BAD2019}: (i) $n_D = 1/2$ and $n_D = - 1/2$ [black and white dots in (b), respectively] indicate half of the topological charge of the vortex in the quadratic field ${\bf V}\cdot {\bf V}$; (ii)  $n_C = 1/2$ and $n_C = - 1/2$ [magenta and cyan dots in (c), respectively] correspond to the number of turns of the major semiaxis of the polarization ellipse along a closed contour including the C-point.
\label{Fig1}}
\end{figure*}

Being thoroughly described and observed for optical fields, polarization singularities and topological polarization structures have not been \red{examined properly} in sound and water waves. 
In this work, we fill this gap. We consider both random and regular structured acoustic and water-surface wavefields and show that polarization singularities and M\"{o}bius strips are also ubiquitous for them.
These results can have a twofold impact. First, they provide a new platform for studying polarization singularities and topological structures. Importantly, while one cannot directly observe elliptical motion of the electric field ${\bm {\mathcal E}}$ or the vector-potential ${\bm {\mathcal A}}$ in optics, the velocity and displacement fields ${\bm {\mathcal V}}$ and ${\bm {\mathcal R}}$ are {\it directly observable} in acoustic and water-surface waves \cite{Taylor1976,Gabrielson1995,surfacewave,Francois2017,Bliokh2020}.
Second, the vector representation of sound and water-surface waves can be relevant for wave-matter interactions, such as interactions with dipole particles coupled to the vector velocity field \cite{Toftul2019,Wei2020,Long2020_II}.

\red{The paper is organized as follows. We start by presenting the generic appearance of C-points and L-lines in random 2D acoustic fields in Section II. Section III presents polarization singularities (C-points) in 3D acoustic fields, both random and regular, as well as polarization M\"{o}bius strips which appear around C-points. Section IV demonstrates the apperance of C-points and polarization M\"{o}bius strips in structured water-surface waves. Concluding remarks are provided in Section V.}

\section{Polarization singularities in a 2D acoustic field}
We consider monochromatic sound waves in a homogeneous fluid or gas, which are described by the equations \cite{LLfluid}:
\begin{equation}
\label{eq1}
i\,\omega\, \beta\, P = {\bm \nabla}  \cdot {\bf{V}}~,\quad
i\,\omega\,\rho\, {\bf V} = {\bm \nabla} P~.
\end{equation}
Here, $\omega$ is the frequency, $\rho$ and $\beta$ are the density and compressibility of the medium, whereas $P({\bf r})$ and ${\bf V}({\bf r})$ are the complex pressure and velocity fields. The real time-dependent fields are ${{\mathcal P}}({\bf r},t) = {\rm Re}[ P({\bf r}) \exp(-i\omega t)]$ and ${\bm {\mathcal V}}({\bf r},t) = {\rm Re}[ {\bf V}({\bf r}) \exp(-i\omega t)]$. 

The plane-wave solution of Eqs.~(\ref{eq1}) is 
\begin{equation}
\label{eq2}
P = P_0 \exp(i\,{\bf k}\cdot{\bf r})~,\quad
{\bf V} = V_0\, \bar{\bf k} \exp(i\,{\bf k}\cdot{\bf r})~,
\end{equation}
%
where $\bar{\bf k} = {\bf k}/k$, ${\bf k}$ is the wavevector, $k = \omega/c$ is the wavenumber, $c=1/\sqrt{\rho\beta}$ is the speed of sound, and $P_0 = \sqrt{\rho/\beta}\, V_0$.
Sound waves are longitudinal because ${\bf V} \parallel {\bf k}$, but still have a vector nature described by the velocity field ${\bf V}$ \cite{Shi2019,Bliokh2019_II,Toftul2019,Rondon2019,Long2020,Burns2020}.
In what follows, we will focus on the polarization properties of this vector wave field: 
the real velocity field ${\bm {\mathcal V}} ({\bf r},t)$ at a given point ${\bf r}$ traces a {\it polarization ellipse}. 

\begin{figure*}[t]
\includegraphics[width=0.9\linewidth]{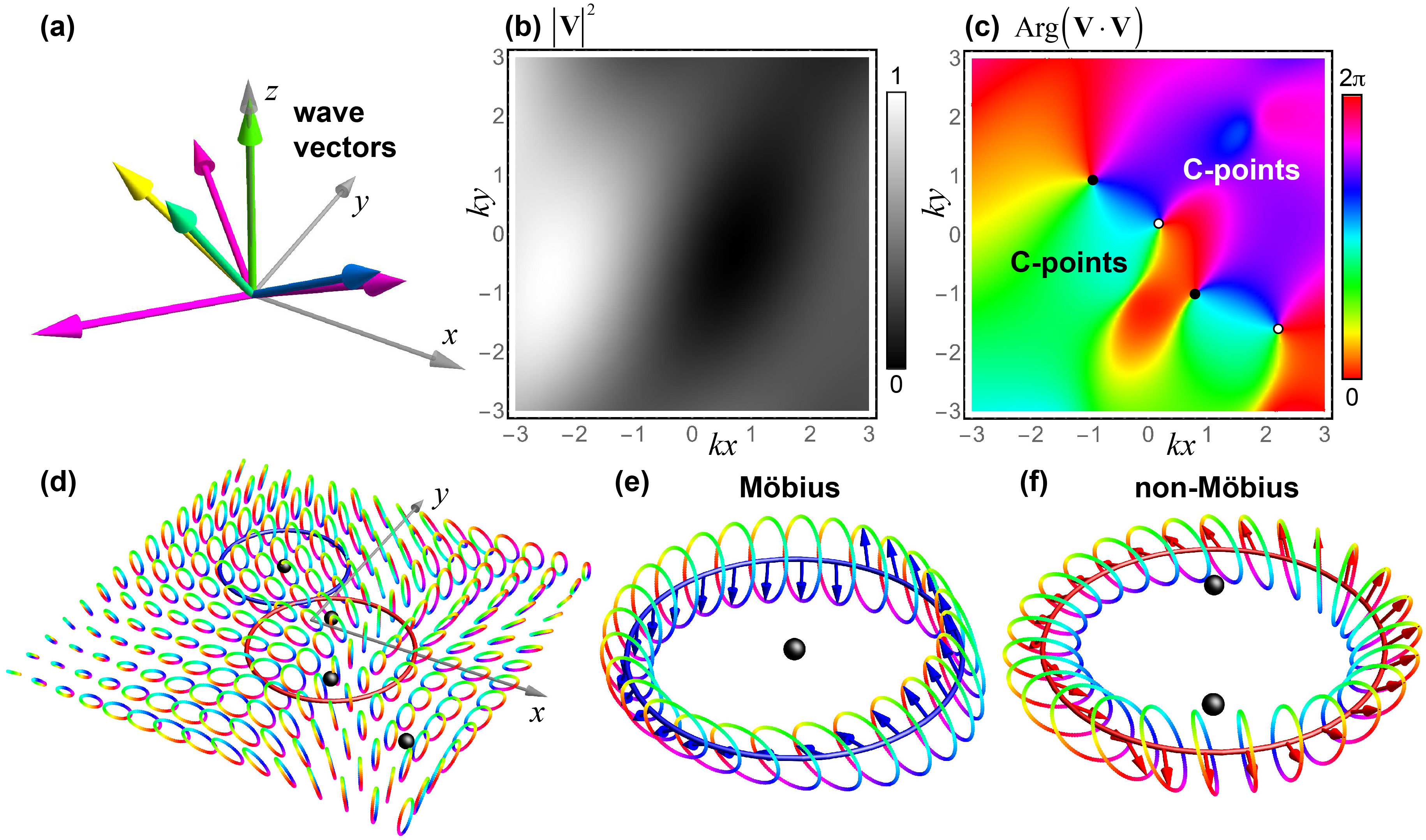}
\caption{Random 3D acoustic field obtained by the interference of $N=7$ plane waves with the same frequency and amplitude but with random directions within the $k_z>0$ semisphere and random phases. (a) Wavevectors of the interfering waves with color-coded phases. (b) Distribution of the intensity of the velocity field $|{\bf V}|^2$ in the $z=0$ plane. (c) Color-coded distribution of the phase of the quadratic field ${\bf V}\cdot {\bf V}$ in the $z=0$ plane. Phase singularities (vortices) of this field correspond to C-points of purely circular polarization in the field ${\bf V}$. (d) Distribution of the normalized polarization ellipses of ${\bf V}$ in the $z=0$ plane. (e) Continuous evolution of polarization ellipses with their major semiaxes along a contour encircling a nondegenerate C-point exhibits a M\"{o}bius-strip structure \cite{Freund2010,Freund2010II,Bauer2015}. (f) For a contour encircling an even number of C-points (including zero), there is no polarization M\"{o}bius strip \cite{BAD2019}. 
\label{Fig2}}
\end{figure*}

We first examine a random speckle-like sound-wave field in 2D. Namely, we consider the interference of $N$ plane waves (\ref{eq2}), ${\bf V} = \sum_{j=1}^{N} {\bf V}_j $, with 
wavevectors ${\bf k}_j$, $j=1,...,N$ randomly distributed over the circle $k_x^2+k_y^2 = k^2$ ($k_z = 0$), and with equal amplitudes $|V_{0j}|$ but random phases $\phi_j = {\rm Arg} (V_{0j})$, as shown in Fig.~\ref{Fig1}(a). Due to the longitudinal character of sound waves, $V_z = 0$, and the polarization ellipses of such field all lie in the $(x,y)$ plane. Figure~\ref{Fig1} shows an example of such random field including its intensity and polarization distributions.    

The distributions in Fig.~\ref{Fig1} are similar to the corresponding distributions in random paraxial electromagnetic fields, with wavevectors directed almost along the $z$-axis and polarization ellipses approximately lying in the $(x,y)$ plane \cite{Nye1983,Dennis2009,BAD2019}. The only difference is that paraxial electromagnetic fields have a typical inhomogeneity scale of $(\theta k)^{-1}$, where $\theta \ll 1$ is the small characteristic angle between the wavevectors and the $z$-axis, while in the acoustic case $\theta = \pi/2$ and the typical inhomogeneity scale is $k^{-1}$. Polarization singularities of generic 2D polarization fields are the {\it C-points} of purely-circular polarization and {\it L-lines} of purely-linear polarization \cite{Nye1987,BerryDennis2001,Nye1983, Hajnal1987, Dennis2009,BAD2019}, as shown in Figs.~\ref{Fig1}(b,c). 

C-points correspond to phase sigularities (vortices) in the scalar field $\Psi({\bf r}) = {\bf V}({\bf r})\cdot {\bf V}({\bf r})$ \cite{BerryDennis2001,Dennis2009,BAD2019},  Fig.~\ref{Fig1}(b). 
Notably, these points generically coincide neither with zeros of the scalar pressure field $P({\bf r})$, nor with zeros of $|{\bf V}({\bf r})|^2$. 
Furthermore, each C-point in a 2D polarization field can be characterized by two half-integer topological numbers \cite{BAD2019}. The first, $n_C$, corresponds to the number of turns of the major semiaxis of the polarization ellipse along a closed contour including the C-point. The second, $n_D$, is half the topological charge of the corresponding phase singularity in the field $\Psi$. In the generic (non-degenerate) case, singularities have the topological numbers $n_C = \pm 1/2$ and $n_D = \pm 1/2$ (see Fig.~\ref{Fig1}). 
\red{Note that the morphological classification of 2D polarization distributions around C-points, such as `stars', `lemons', and `monstars', is thoroughly described for optical polarized fields \cite{mrd2002,Dennis2009} and applies here as well.} 
Note also that higher-order singularities can appear in degenerate cases, e.g., with imposed additional symmetries, such as cylindrical beams.   


\section{C-points and polarization M\"{o}bius strips in 3D acoustic fields}
\subsection{Random fields}
Akin to nonparaxial 3D electromagnetic fields, generic sound-wave fields have polarization characterized by the ellipses traced by the velocity field ${\bm {\mathcal V}} ({\bf r},t)$ at every point ${\bf r}$, which can be arbitrarily oriented in 3D space. To show an example of such field, we consider an interference of $N$ plane waves with equal amplitudes $|V_{0j}|$, wavevectors ${\bf k}_j$, $j=1,...,N$, with directions randomly distributed over the hemisphere $k_z > 0$ ($k_x^2+k_y^2+k_z^2 = k^2$), and random phases $\phi_j = {\rm Arg} (V_{0j})$, see Fig.~\ref{Fig2}(a).

\begin{figure*}[t]
\includegraphics[width=0.8\linewidth]{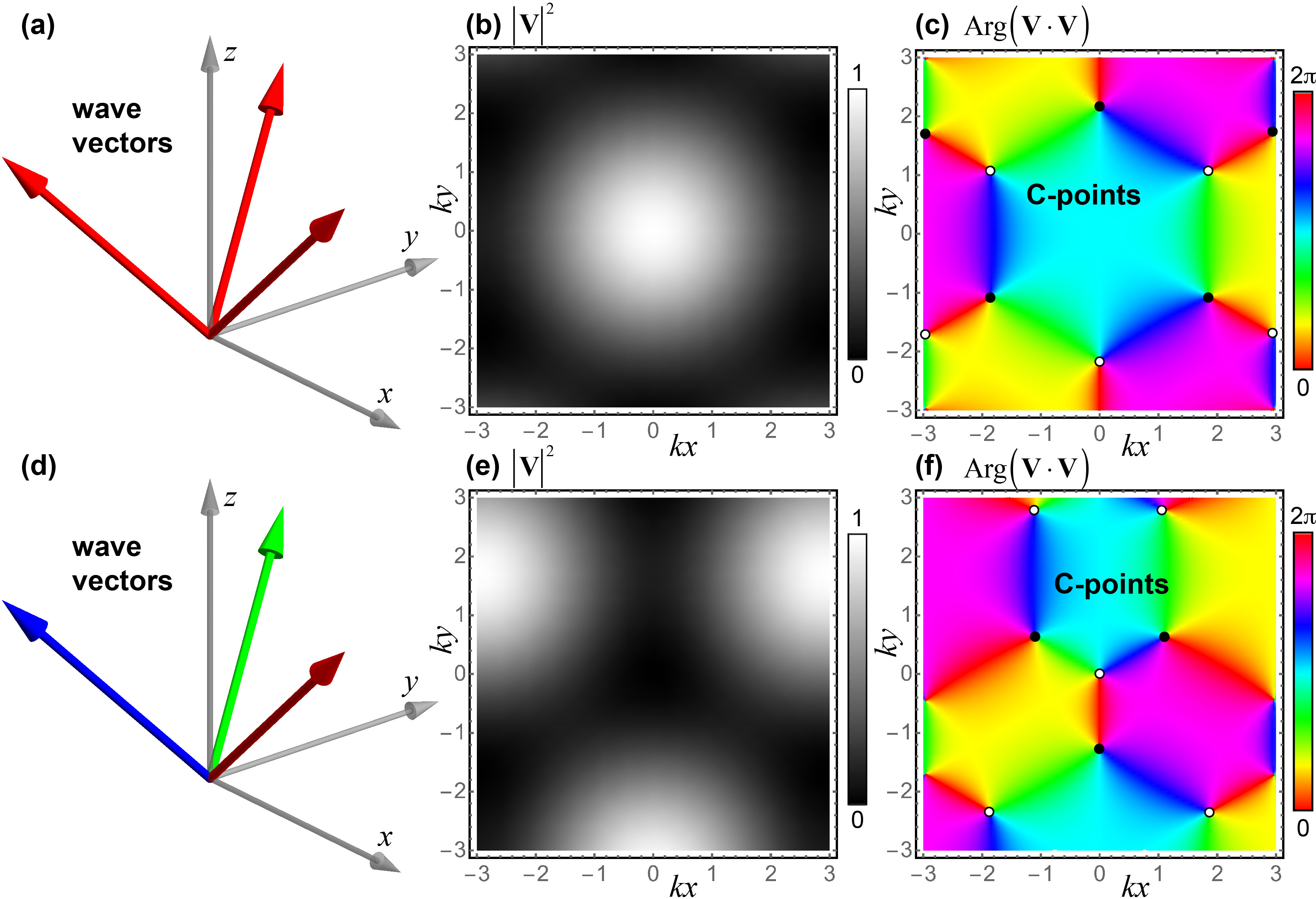}
\caption{The wavevectors ${\bf k}_j$, distributions of the intensity of the velocity field, $|{\bf V}|^2$, and of the phase of the quadratic field ${\bf V}\cdot{\bf V}$ in the $z=0$ plane, for the three-wave superpositions (\ref{eq3}) with $N=3$, $\theta_0 = \pi/4$, $\ell=0$ (a,b,c) and $\ell = 1$ (d,e,f).  
\label{Fig3}}
\end{figure*}

The distributions of the resulting intensity $|{\bf V}|^2$ and of the phase of the quadratic field ${\bf V}\cdot {\bf V}$ over the $z=0$ plane are shown in Figs.~\ref{Fig2}(b,c). Similar to the 2D case, the phase singularities of the quadratic field correspond to the C-points (polarization singularities) in the polarization distribution, Fig.~\ref{Fig2}(d). However,  in the 3D case, the circular polarizations in these points do not generically lie in the $(x,y)$ plane \cite{Nye1987,BerryDennis2001,Dennis2009,BAD2019}. 

Furthermore, distributions of the 3D polarization ellipses in the vicinity of C-points have remarkable topological properties. Namely, continuous evolution of the major semiaxes of the polarization ellipse along a contour encircling a non-degenerate C-point traces a 3D {\it M\"{o}bius-strip}-like structure \cite{Freund2010,Freund2010II,Bauer2015,Galvez2017,Garcia2017,Kreismann2017, Bauer2019, Tekce2019, BAD2019}, Fig.~\ref{Fig2}(e). Notably, the number of turns of the polarization ellipse around the contour is not topologically stable: continuous deformations of the contour (without crossing C-points) can result in the change of the number of turns by an integer number \cite{Dennis2011,Freund2020}. However, the number of turns modulo 1/2, which distinguish the `M\"{o}bius' (half-integer number of turns) and `non-M\"{o}bius' (integer number of turns) cases is topologically stable. It directly corresponds to the number of C-points enclosed by the contour modulo 2 \cite{BAD2019}, see Figs.~\ref{Fig2}(d,e,f).  

Recently, polarization M\"{o}bius strips attracted great attention in optics \cite{Bauer2015,Galvez2017,Garcia2017,Kreismann2017, Bauer2019, Tekce2019, BAD2019}. We argue that entirely similar polarization structures naturally appear in inhomogeneous sound-wave fields. In addition to the random field shown in Fig.~\ref{Fig2}, below we consider examples of regular sound-wave fields with polarization singularities and M\"{o}bius strip. 

\subsection{Three-wave interference}
We now consider examples of regular (non-random) 3D acoustic fields with polarization singularities and M\"{o}bius strips. In optics, such singularities are often generated in vector vortex beams \cite{Dennis2009,Bauer2015,Galvez2017,Bauer2019,Tekce2019,BAD2019}. Here we also consider a superposition of $N$ acoustic plane waves with wavevectors evenly distributed within a cone of polar angle $\theta=\theta_0$ and with an azimuthal phase difference corresponding to a vortex of order $\ell$: 
\begin{equation}
\label{eq3}
{\bf V} = V_0\, \sum_{j=1}^{N}\bar{\bf k}_j \exp\! \left[ i\,{\bf k}_j\cdot {\bf r} + i\,\ell\,\phi_j \right]~,
\end{equation}
where ${\bf k}_j = k \left( \sin \theta_0 \cos \phi_j, \,\sin \theta_0 \sin \phi_j, \,\cos \theta_0 \right)$ and $\phi_j = 2\pi(j-1)/N$.
In the limit of $N\gg 1$, this superposition tends to an acoustic Bessel beam \cite{Bliokh2019_II}.

\begin{figure*}[t]
\includegraphics[width=0.8\linewidth]{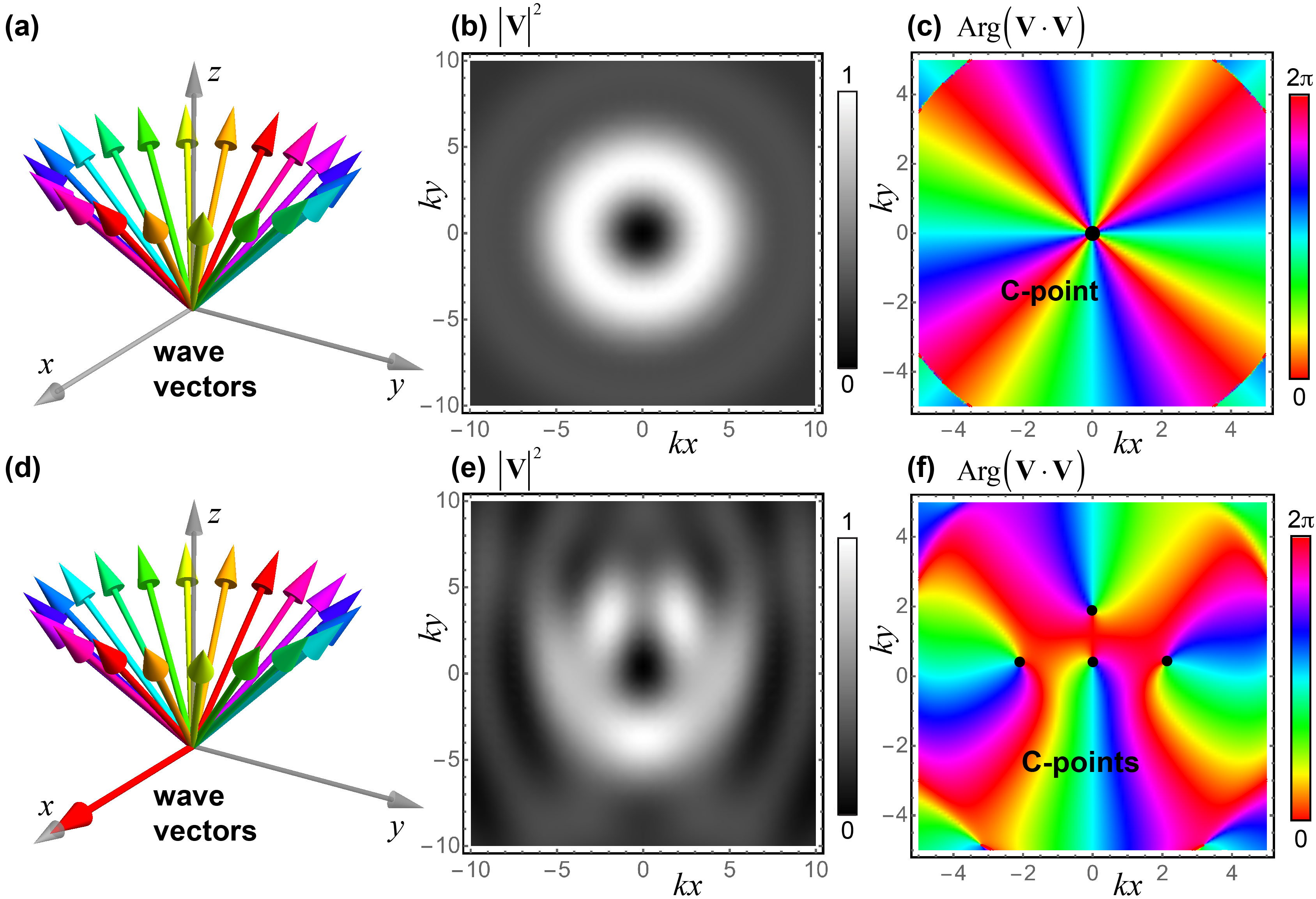}
\caption{The wavevectors, distributions of the intensity of the velocity field, $|{\bf V}|^2$, and of the phase of the quadratic field ${\bf V}\cdot{\bf V}$ in the $z=0$ plane, for the Bessel-beam superposition (\ref{eq3}) with $N=20$, $\theta_0 = \pi/4$, $\ell=2$ (a,b,c) and the same Bessel beam interfering with an additional plane wave, Eq.~(\ref{eq4}), with $V'/V_0 = 2$ (d,e,f). One can see the splitting of the even-order C-point (integer $n_D = \ell$) (c) into $2|n_D|$ first-order C-points (f) when breaking the cylindrical symmetry of the vortex beam.  
\label{Fig4}}
\end{figure*}

The minimal number of plane waves to generate polarization singularities is $N=3$. Figure~\ref{Fig3} shows the wavevectors ${\bf k}_j$, distributions of the intensity of the velocity field, $|{\bf V}|^2$, and of the phase of the quadratic field, ${\rm Arg}({\bf V}\cdot{\bf V})$, for the three-wave superpositions (\ref{eq3}) with $\theta_0 = \pi/4$, $\ell=0$ and $\ell = 1$. One can see a number of first-order C-points, i.e., phase singularities in the quadratic field ${\bf V}\cdot{\bf V}$. Accordingly, 3D polarization ellipses along a contour enclosing an odd number of C-points form polarization M\"{o}bius strips. Importantly, the spacing between the C-points in Fig.~\ref{Fig3} is controlled by the polar angle $\theta_0$. When $\theta_0 \ll 1$ (paraxial regime), the C-points merge and form only {\it even-order} C-points with no M\"{o}bius strips around them. 
In particular, the four C-points at the center of Fig.~\ref{Fig3}(f) with the integer total topological charge $n_D = 3/2 -1/2 =1$ merge into a single second-order ($n_D = 1$) C-point in the paraxial limit. This is in sharp contrast to the electromagnetic (optical) waves, where isolated first-order C-points can appear even in the paraxial case.

\subsection{\red{Perturbed} vortex beams}
Consider now the large-$N$ limit of the superposition (\ref{eq3}), which generates acoustic vortex (Bessel) beams. Due to the cylindrical symmetry, such beams can have an isolated C-point at the center. However, in contrast to optical vectorial vortex beams, the C-point at the center of an acoustic vortex beam always has an even order (integer $n_D = \ell$) \cite{Bliokh2019_II} (see Fig.~\ref{Fig4}). This does not allow one to generate an acoustic polarization M\"{o}bius strip in a symmetric vortex configuration as in optics \cite{Bauer2015,Galvez2017,Bauer2019,Tekce2019,BAD2019}. However, breaking the cylindrical symmetry of the beam results in splitting of the even-order C-point at the center into a number of the first-order C-points ($n_D = \pm 1/2$), each of which carries polarization M\"{o}bius strip structures around it. For example, one can break the symmetry by interfering the vortex beam with a horizontally-propagating plane wave: 
\begin{equation}
\label{eq4}
{\bf V} = {\bf V}_{\rm vortex} + V' \bar{\bf k}' \exp(i\,{\bf k}'\cdot {\bf r})~,
\end{equation}
where ${\bf V}_{\rm vortex}$ is the vortex-beam field, such as Eq.~(\ref{eq3}) with $N\gg 1$, whereas ${\bf k}' =k(1,0,0)$. Figure~\ref{Fig4} shows the splitting of the even-order C-point at the center of an acoustic vortex beam into first-order C-points when interfering the beam with a horizontally-propagating plane wave.

The above examples show that the typical spacing between the C-points in structured sound waves is $k^{-1}$, and this spacing can decrease in the paraxial regime and in the presence of additional symmetries. This also determines the typical subwavelength size of the acoustic polarization M\"{o}bius strips.  

\begin{figure*}
\includegraphics[width=0.75\linewidth]{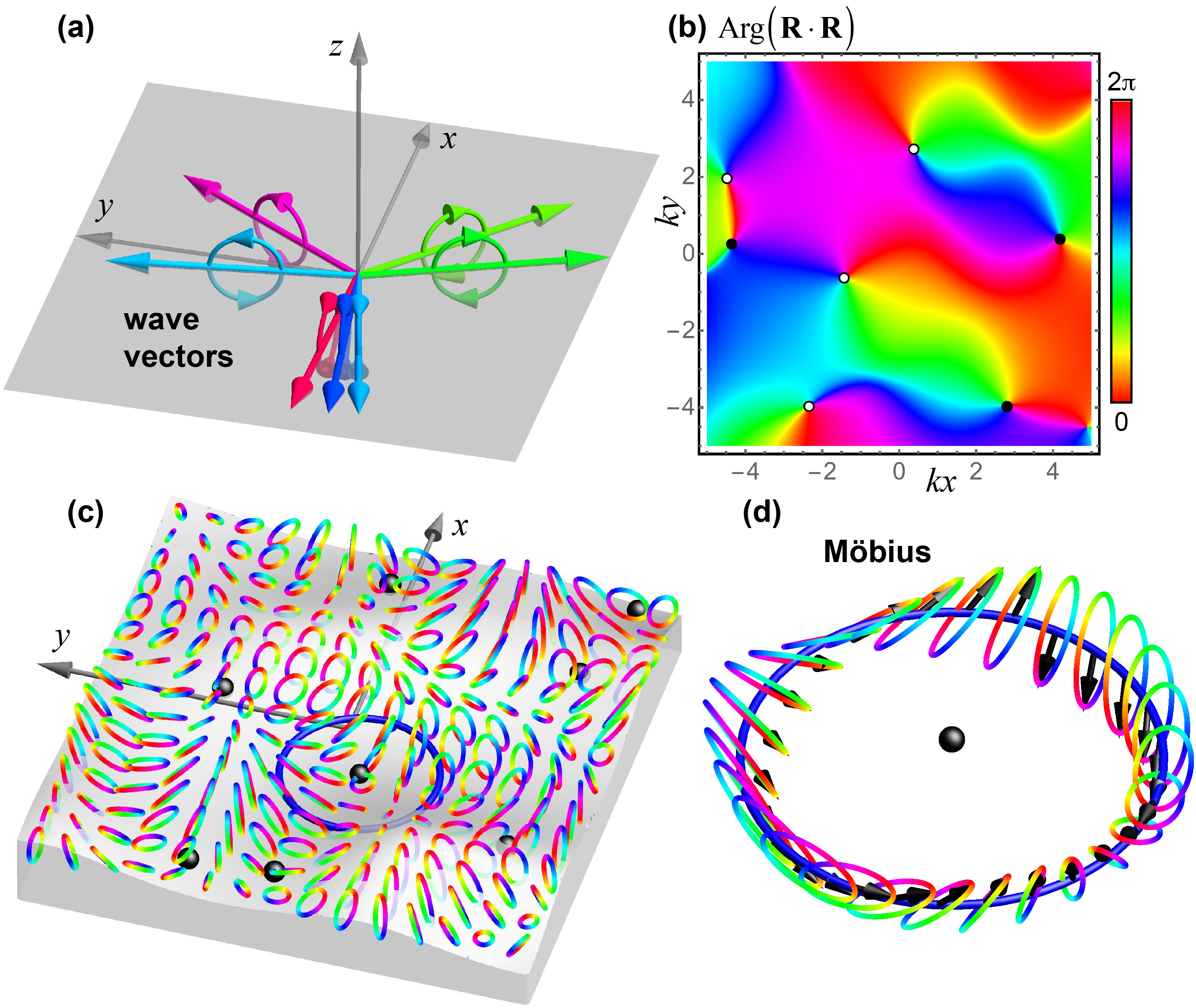}
\caption{
Random 3D water-surface wave field obtained by the interference of $N=7$ plane waves with the same frequency and amplitude but with random directions in the $(x,y)$ plane and random phases. (a) Wavevectors of the interfering waves with color-coded phases and their circular polarizations. (b) Color-coded distribution of the phase of the quadratic field ${\bf R}\cdot {\bf R}$ in the water-surface $z=0$ plane. Phase singularities (vortices) of this field correspond to C-points of purely circular polarization in the field ${\bf R}$. (c) Distribution of the normalized polarization ellipses of the ${\bf R}$ field in the $z=0$ plane. These ellipses are trajectories of water-surface particles (Multimedia view). The instantaneous water surface at $t=0$ is shown in gray. (d) Akin to Fig.~\ref{Fig2}, the continuous evolution of polarization ellipses with their major semiaxes along a contour encircling a C-point exhibits a polarization M\"{o}bius strip (Multimedia view).
\label{Fig5}}
\end{figure*}

\section{Polarization singularities and M\"{o}bius strips in water-surface waves}
One of the key differences between electromagnetic and acoustic fields is that the electric and magnetic fields are vectors in abstract spaces of the field components (there is no `ether' and nothing moves in a free-space electromagnetic wave), while the velocity field corresponds to the motion of the medium's particles (atoms or molecules) in real space. Moreover, instead of the velocity field, one can consider the {\it displacement} field ${\bm {\mathcal R}}({\bf r},t)$: ${\bm {\mathcal V}} = \partial_t {\bm {\mathcal R}}$, or, for a monochromatic field in the complex representation, ${\bf V} = -i\,\omega\, {\bf R}$. The displacement field can be regarded as a `vector-potential' for the velocity field \cite{Burns2020}. It has the same polarization, but the polarization ellipses traced by ${\bm {\mathcal R}}$ correspond to {\it real-space trajectories} of the medium particles. 

This opens an avenue to the {\it direct observation} of polarization ellipses and more complicated structures \cite{Sugic2020}.
 In sound waves, typical displacement amplitudes are small and their direct observations are challenging \cite{Taylor1976,Gabrielson1995}. However, similar medium displacements can be easily observed in another type of classical waves, namely, {\it water-surface (e.g., gravity or capillary) waves} \cite{LLfluid}, with typical displacement scales ranging from milimeters to meters. Recently, there were several studies on polarization properties of structured water waves \cite{Francois2017,Bliokh2020,Sugic2020}, and here we show that these waves naturally reveal generic polarization singularities.

For the sake of simplicity, we consider deep-water gravity waves on the unperturbed water surface $z=0$ \cite{LLfluid}. The equations of motion for the complex displacement field ${\bf R} = (X,Y,Z)$ of the water-surface particles in a monochromatic wave field can be written in a form similar to the acoustic equations (\ref{eq1}) \cite{Bliokh2020,Sugic2020}:
\begin{equation}
\label{eq5}
\omega^2 Z = -g\, {\bm \nabla}_\perp  \cdot {\bf R}_\perp~,\quad
\omega^2 {\bf R}_\perp = g\, {\bm \nabla}_\perp Z~.
\end{equation}
Here $g$ is the gravitational acceleration, ${\bf R}_\perp = (X,Y)$, and ${\bm \nabla}_\perp = (\partial_x,\partial_y)$. Making the plane-wave ansatz ${\bm \nabla}_\perp \to i\, {\bf k}$, ${\bf k} = (k_x,k_y)$, in Eqs.~(\ref{eq5}), we obtain the dispersion relation $\omega^2 = gk$.

The plane-wave solution of Eqs.~(\ref{eq5}) is:
\begin{equation}
\label{eq6}
Z = Z_0 \exp(i\,{\bf k}\cdot{\bf r})~,\quad
{\bf R}_\perp = i\, Z_0\, \bar{\bf k} \exp(i\,{\bf k}\cdot{\bf r})~.
\end{equation}
These relations show that deep-water gravity waves have equal longitudinal (${\bf k}$-directed) and transverse ($z$-directed) displacement components phase-shifted by $\pi/2$ with respect to each other. In other words, such plane waves are {\it circularly polarized} in the meridional (propagation) plane including the wavevector and the normal to the unperturbed water surface \cite{surfacewave}. 
Such (generically, elliptical) meridional polarization is a common feature of surface and evanescent waves in different physical contexts \cite{Shi2019,Bliokh2019,Bliokh2015PR,Aiello2015}.
Therefore, interfering plane water waves with wavevectors lying in the $(x,y)$ plane results in generic 3D polarization structures with all three components of the displacement field ${\bf R}$.

To show that such polarization distributions generically posses polarization singularities, we consider the interference of $N$ plane waves (\ref{eq6}): ${\bf R} = \sum_{j=1}^{N} {\bf R}_j \equiv \sum_{j=1}^{N} {\bf R}_{0j} \exp(i\,{\bf k}_j \cdot {\bf r}_\perp)$, with ${\bf R}_{0j} = Z_{0j} (i\bar{k}_{jx},i\bar{k}_{jy},1) $,  
wavevectors ${\bf k}_j$ randomly distributed over the circle $k_x^2+k_y^2 = k^2$,  and equal amplitudes $|Z_{0j}|$ but random phases $\phi_j = {\rm Arg} (Z_{0j})$, as shown in Fig.~\ref{Fig5}(a) (cf. Fig.~\ref{Fig1}). Figure~\ref{Fig5}(b) shows the phase distribution of the quadratic field $\Psi = {\bf R}\cdot {\bf R}$; it clearly exhibits phase singularities corresponding to C-points of the vector field ${\bf R}$. The distribution of the polarization ellipses, i.e., trajectories of the water-surface particles, over the $z=0$ plane is shown in Fig.~\ref{Fig5}(c) (Multimedia view). Tracing the orientation of the polarization ellipses along a contour encircling a C-point reveals the generic M\"{o}bius-strip structure, Fig.~\ref{Fig5}(d) (Multimedia view).
We also provide animated versions of Figs.~\ref{Fig5}(c,d), where one can see motion of the water surface and separate water particles. 
In particular, the animated version of Fig.~\ref{Fig5}(d) shows the temporal evolution of the displacement vectors $\bm{\mathcal R} ({\bf r},t)$ along the contour, which can form `twisted ribbon carousels' \cite{Freund2020_II}.

Thus, by tracing 3D trajectories of water particles in a random (yet monochromatic) water-surface wavefield one can directly observe generic polarization singularities of 3D vector wavefields. 

\section{Concluding remarks}

This work was motivated by recent strong interest in (i) polarization M\"{o}bius strips in 3D polarized optical fields \cite{Freund2010,Freund2010II,Dennis2011,Bauer2015,Galvez2017,Garcia2017,Kreismann2017, Bauer2019, Tekce2019, BAD2019}
and (ii) vectorial spin properties of acoustic and water-surface waves 
\cite{Shi2019,Bliokh2019,Bliokh2019_II,Toftul2019,Rondon2019,Long2020, Burns2020,Bliokh2020,Sugic2020}.
We have shown that these research directions can be naturally coupled, and that polarization singularities, such as C-points and polarization M\"{o}bius strips, are ubiquitous for inhomogeneous (yet monochromatic) acoustic and water-surface waves. The vector velocity or displacement of the medium particles provide complex-valued elliptical polarization fields varying across the space. We have considered various examples of random and regular interference fields consisting of multiple (three or more) plane waves, which exhibit polarization singularities and M\"{o}bius strips. 

In contrast to well-studied electromagnetic polarizations associated with the motion of abstract field vectors, acoustic and water-wave polarizations correspond to real-space trajectories of the medium particles. In particular, these are readily directly observable for water-surface waves \cite{Francois2017,Bliokh2020}. Also, while optical vectorial-vortex beams can bear an isolated first-order C-point and a M\"{o}bius strip around it \cite{Bauer2015,Galvez2017,Bauer2019,Tekce2019,BAD2019}, acoustic C-points typically appear in clusters with subwavelength distance between the points.

Analyzing wave-field singularities is useful because of their topological robustness; they provide a `skeleton' of an inhomogeneous field \cite{Nye_book}. 
\red{This robustness is highly important because real-life waves in fluids always have inherent perturbations, such as viscosity and nonlinearity, with respect to the idealized non-dissipative linear picture.}
So far, only {\it phase} singularities of the {\it scalar} pressure field $P$ were considered in sound-wave fields. The vector velocity field ${\bf V} \propto \bm{\nabla} P$ and its polarization singularities provide an alternative representation and can be more relevant, e.g., in problems involving dipole wave-matter coupling. Note that the vectorial representation of a gradient of a scalar wavefield was previously considered in Ref.~\cite{Dennis2004_II}. 

For water-surface waves, the scalar representation is based on the vertical displacement field $Z$. Tidal ocean waves were also studied in terms of the 2D polarization field of the horizontal current \cite{Hansen,Berry2001,Nye1988,Ray2001,Ray2004} associated with the velocity components $(V_x, V_y)$. We argue that these scalar and 2D vector fields can be regarded as components of a single 3D vector displacement ${\bf R}$ or velocity ${\bf V}$ field. Moreover, we have considered gravity deep-water waves, which are much more feasible for experimental laboratory studies than tidal waves \cite{Francois2017,Bliokh2020}.  

Notably, our arguments are not restricted to purely sound and water-surface waves. They can be equally applied to any fluid/gas or fluid/fluid surface waves as well as internal gravity waves in stratified fluid or gas media. 
We hope that our work will stimulate further studies and possibly applications of 3D polarization textures and topological vectorial properties of various waves in acoustics and fluid mechanics. 

\begin{acknowledgments}
M.A.A. was supported by the Excellence Initiative of Aix Marseille University---A*MIDEX, a French `Investissements d'Avenir' programme.
F.N. was supported by 
Nippon Telegraph and Telephone Corporation (NTT) Research; 
the Japan Science and Technology Agency (JST) via the Quantum Leap Flagship Program (Q-LEAP), the Moonshot R\&D Grant No. JP- MJMS2061, and the Centers of Research Excellence in Science and Technology (CREST) Grant No. JPMJCR1676; 
the Japan Society for the Promotion of Science (JSPS) via the Grants-in-Aid for Scientific Research (KAKENHI) Grant No. JP20H00134, and the JSPS–RFBR Grant No. JPJSBP120194828; 
the Army Research Office (ARO) (Grant No. W911NF-18-1-0358); 
the Asian Office of Aerospace Research and Development (AOARD) (Grant No. FA2386-20-1-4069); 
and the Foundational Questions Institute Fund (FQXi) (Grant No. FQXi-IAF19-06).  
\end{acknowledgments}



%

\bibliography{References_revised.bib}

\begin{thebibliography}{56}%
\makeatletter
\providecommand \@ifxundefined [1]{%
 \@ifx{#1\undefined}
}%
\providecommand \@ifnum [1]{%
 \ifnum #1\expandafter \@firstoftwo
 \else \expandafter \@secondoftwo
 \fi
}%
\providecommand \@ifx [1]{%
 \ifx #1\expandafter \@firstoftwo
 \else \expandafter \@secondoftwo
 \fi
}%
\providecommand \natexlab [1]{#1}%
\providecommand \enquote  [1]{``#1''}%
\providecommand \bibnamefont  [1]{#1}%
\providecommand \bibfnamefont [1]{#1}%
\providecommand \citenamefont [1]{#1}%
\providecommand \href@noop [0]{\@secondoftwo}%
\providecommand \href [0]{\begingroup \@sanitize@url \@href}%
\providecommand \@href[1]{\@@startlink{#1}\@@href}%
\providecommand \@@href[1]{\endgroup#1\@@endlink}%
\providecommand \@sanitize@url [0]{\catcode `\\12\catcode `\$12\catcode
  `\&12\catcode `\#12\catcode `\^12\catcode `\_12\catcode `\%12\relax}%
\providecommand \@@startlink[1]{}%
\providecommand \@@endlink[0]{}%
\providecommand \url  [0]{\begingroup\@sanitize@url \@url }%
\providecommand \@url [1]{\endgroup\@href {#1}{\urlprefix }}%
\providecommand \urlprefix  [0]{URL }%
\providecommand \Eprint [0]{\href }%
\providecommand \doibase [0]{http://dx.doi.org/}%
\providecommand \selectlanguage [0]{\@gobble}%
\providecommand \bibinfo  [0]{\@secondoftwo}%
\providecommand \bibfield  [0]{\@secondoftwo}%
\providecommand \translation [1]{[#1]}%
\providecommand \BibitemOpen [0]{}%
\providecommand \bibitemStop [0]{}%
\providecommand \bibitemNoStop [0]{.\EOS\space}%
\providecommand \EOS [0]{\spacefactor3000\relax}%
\providecommand \BibitemShut  [1]{\csname bibitem#1\endcsname}%
\let\auto@bib@innerbib\@empty
\bibitem [{\citenamefont {Azzam}\ and\ \citenamefont
  {Bashara}(1977)}]{Azzam_book}%
  \BibitemOpen
  \bibfield  {author} {\bibinfo {author} {\bibfnamefont {R.~M.~A.}\
  \bibnamefont {Azzam}}\ and\ \bibinfo {author} {\bibfnamefont {N.~M.}\
  \bibnamefont {Bashara}},\ }\href@noop {} {\emph {\bibinfo {title}
  {Ellipsometry and polarized light}}}\ (\bibinfo  {publisher}
  {North-Holland},\ \bibinfo {year} {1977})\BibitemShut {NoStop}%
\bibitem [{\citenamefont {Andrews}\ and\ \citenamefont
  {Babiker}(2012)}]{Andrews_book}%
  \BibitemOpen
  \bibinfo {editor} {\bibfnamefont {D.~L.}\ \bibnamefont {Andrews}}\ and\
  \bibinfo {editor} {\bibfnamefont {M.}~\bibnamefont {Babiker}},\ eds.,\
  \href@noop {} {\emph {\bibinfo {title} {{The Angular Momentum of Light}}}}\
  (\bibinfo  {publisher} {Cambridge University Press},\ \bibinfo {year}
  {2012})\BibitemShut {NoStop}%
\bibitem [{\citenamefont {Berestetskii}\ \emph {et~al.}(1982)\citenamefont
  {Berestetskii}, \citenamefont {Lifshitz},\ and\ \citenamefont
  {Pitaevskii}}]{BLP}%
  \BibitemOpen
  \bibfield  {author} {\bibinfo {author} {\bibfnamefont {V.~B.}\ \bibnamefont
  {Berestetskii}}, \bibinfo {author} {\bibfnamefont {E.~M.}\ \bibnamefont
  {Lifshitz}}, \ and\ \bibinfo {author} {\bibfnamefont {L.~P.}\ \bibnamefont
  {Pitaevskii}},\ }\href@noop {} {\emph {\bibinfo {title} {{Quantum
  Electrodynamics}}}}\ (\bibinfo  {publisher} {Pergamon Press, Oxford},\
  \bibinfo {year} {1982})\BibitemShut {NoStop}%
\bibitem [{\citenamefont {Shi}\ \emph {et~al.}(2019)\citenamefont {Shi},
  \citenamefont {Zhao}, \citenamefont {Long}, \citenamefont {Yang},
  \citenamefont {Wang}, \citenamefont {Chen}, \citenamefont {Ren},\ and\
  \citenamefont {Zhang}}]{Shi2019}%
  \BibitemOpen
  \bibfield  {author} {\bibinfo {author} {\bibfnamefont {C.}~\bibnamefont
  {Shi}}, \bibinfo {author} {\bibfnamefont {R.}~\bibnamefont {Zhao}}, \bibinfo
  {author} {\bibfnamefont {Y.}~\bibnamefont {Long}}, \bibinfo {author}
  {\bibfnamefont {S.}~\bibnamefont {Yang}}, \bibinfo {author} {\bibfnamefont
  {Y.}~\bibnamefont {Wang}}, \bibinfo {author} {\bibfnamefont {H.}~\bibnamefont
  {Chen}}, \bibinfo {author} {\bibfnamefont {J.}~\bibnamefont {Ren}}, \ and\
  \bibinfo {author} {\bibfnamefont {X.}~\bibnamefont {Zhang}},\ }\bibfield
  {title} {\enquote {\bibinfo {title} {Observation of acoustic spin},}\
  }\href@noop {} {\bibfield  {journal} {\bibinfo  {journal} {Natl. Sci. Rev.}\
  }\textbf {\bibinfo {volume} {6}},\ \bibinfo {pages} {707} (\bibinfo {year}
  {2019})}\BibitemShut {NoStop}%
\bibitem [{\citenamefont {Bliokh}\ and\ \citenamefont
  {Nori}(2019{\natexlab{a}})}]{Bliokh2019}%
  \BibitemOpen
  \bibfield  {author} {\bibinfo {author} {\bibfnamefont {K.~Y.}\ \bibnamefont
  {Bliokh}}\ and\ \bibinfo {author} {\bibfnamefont {F.}~\bibnamefont {Nori}},\
  }\bibfield  {title} {\enquote {\bibinfo {title} {Transverse spin and surface
  waves in acoustic metamaterials},}\ }\href@noop {} {\bibfield  {journal}
  {\bibinfo  {journal} {Phys. Rev. B}\ }\textbf {\bibinfo {volume} {99}},\
  \bibinfo {pages} {020301(R)} (\bibinfo {year}
  {2019}{\natexlab{a}})}\BibitemShut {NoStop}%
\bibitem [{\citenamefont {Bliokh}\ and\ \citenamefont
  {Nori}(2019{\natexlab{b}})}]{Bliokh2019_II}%
  \BibitemOpen
  \bibfield  {author} {\bibinfo {author} {\bibfnamefont {K.~Y.}\ \bibnamefont
  {Bliokh}}\ and\ \bibinfo {author} {\bibfnamefont {F.}~\bibnamefont {Nori}},\
  }\bibfield  {title} {\enquote {\bibinfo {title} {Spin and orbital angular
  momenta of acoustic beams},}\ }\href@noop {} {\bibfield  {journal} {\bibinfo
  {journal} {Phys. Rev. B}\ }\textbf {\bibinfo {volume} {99}},\ \bibinfo
  {pages} {174310} (\bibinfo {year} {2019}{\natexlab{b}})}\BibitemShut
  {NoStop}%
\bibitem [{\citenamefont {Toftul}\ \emph {et~al.}(2019)\citenamefont {Toftul},
  \citenamefont {Bliokh}, \citenamefont {Petrov},\ and\ \citenamefont
  {Nori}}]{Toftul2019}%
  \BibitemOpen
  \bibfield  {author} {\bibinfo {author} {\bibfnamefont {I.~D.}\ \bibnamefont
  {Toftul}}, \bibinfo {author} {\bibfnamefont {K.~Y.}\ \bibnamefont {Bliokh}},
  \bibinfo {author} {\bibfnamefont {M.~I.}\ \bibnamefont {Petrov}}, \ and\
  \bibinfo {author} {\bibfnamefont {F.}~\bibnamefont {Nori}},\ }\bibfield
  {title} {\enquote {\bibinfo {title} {Acoustic radiation force and torque on
  small particles as measures of the canonical momentum and spin densities},}\
  }\href@noop {} {\bibfield  {journal} {\bibinfo  {journal} {Phys. Rev. Lett.}\
  }\textbf {\bibinfo {volume} {123}},\ \bibinfo {pages} {183901} (\bibinfo
  {year} {2019})}\BibitemShut {NoStop}%
\bibitem [{\citenamefont {Rond\'{o}n}\ and\ \citenamefont
  {Leykam}(2019)}]{Rondon2019}%
  \BibitemOpen
  \bibfield  {author} {\bibinfo {author} {\bibfnamefont {I.}~\bibnamefont
  {Rond\'{o}n}}\ and\ \bibinfo {author} {\bibfnamefont {D.}~\bibnamefont
  {Leykam}},\ }\bibfield  {title} {\enquote {\bibinfo {title} {Acoustic vortex
  beams in synthetic magnetic fields},}\ }\href@noop {} {\bibfield  {journal}
  {\bibinfo  {journal} {J. Phys.: Cond. Mat.}\ }\textbf {\bibinfo {volume}
  {32}},\ \bibinfo {pages} {104001} (\bibinfo {year} {2019})}\BibitemShut
  {NoStop}%
\bibitem [{\citenamefont {Long}\ \emph
  {et~al.}(2020{\natexlab{a}})\citenamefont {Long}, \citenamefont {Zhang},
  \citenamefont {Yang}, \citenamefont {Ge}, \citenamefont {Chen},\ and\
  \citenamefont {Ren}}]{Long2020}%
  \BibitemOpen
  \bibfield  {author} {\bibinfo {author} {\bibfnamefont {Y.}~\bibnamefont
  {Long}}, \bibinfo {author} {\bibfnamefont {D.}~\bibnamefont {Zhang}},
  \bibinfo {author} {\bibfnamefont {C.}~\bibnamefont {Yang}}, \bibinfo {author}
  {\bibfnamefont {J.}~\bibnamefont {Ge}}, \bibinfo {author} {\bibfnamefont
  {H.}~\bibnamefont {Chen}}, \ and\ \bibinfo {author} {\bibfnamefont
  {J.}~\bibnamefont {Ren}},\ }\bibfield  {title} {\enquote {\bibinfo {title}
  {Realization of acoustic spin transport in metasurface waveguides},}\
  }\href@noop {} {\bibfield  {journal} {\bibinfo  {journal} {Nat. Commun.}\
  }\textbf {\bibinfo {volume} {11}},\ \bibinfo {pages} {4716} (\bibinfo {year}
  {2020}{\natexlab{a}})}\BibitemShut {NoStop}%
\bibitem [{\citenamefont {Burns}\ \emph {et~al.}(2020)\citenamefont {Burns},
  \citenamefont {Bliokh}, \citenamefont {Nori},\ and\ \citenamefont
  {Dressel}}]{Burns2020}%
  \BibitemOpen
  \bibfield  {author} {\bibinfo {author} {\bibfnamefont {L.}~\bibnamefont
  {Burns}}, \bibinfo {author} {\bibfnamefont {K.~Y.}\ \bibnamefont {Bliokh}},
  \bibinfo {author} {\bibfnamefont {F.}~\bibnamefont {Nori}}, \ and\ \bibinfo
  {author} {\bibfnamefont {J.}~\bibnamefont {Dressel}},\ }\bibfield  {title}
  {\enquote {\bibinfo {title} {Acoustic versus electromagnetic field theory:
  scalar, vector, spinor representations and the emergence of acoustic spin},}\
  }\href@noop {} {\bibfield  {journal} {\bibinfo  {journal} {New J. Phys.}\
  }\textbf {\bibinfo {volume} {22}},\ \bibinfo {pages} {053050} (\bibinfo
  {year} {2020})}\BibitemShut {NoStop}%
\bibitem [{\citenamefont {Bliokh}\ \emph {et~al.}(2020)\citenamefont {Bliokh},
  \citenamefont {Punzmann}, \citenamefont {Xia}, \citenamefont {Nori},\ and\
  \citenamefont {Shats}}]{Bliokh2020}%
  \BibitemOpen
  \bibfield  {author} {\bibinfo {author} {\bibfnamefont {K.~Y.}\ \bibnamefont
  {Bliokh}}, \bibinfo {author} {\bibfnamefont {H.}~\bibnamefont {Punzmann}},
  \bibinfo {author} {\bibfnamefont {H.}~\bibnamefont {Xia}}, \bibinfo {author}
  {\bibfnamefont {F.}~\bibnamefont {Nori}}, \ and\ \bibinfo {author}
  {\bibfnamefont {M.}~\bibnamefont {Shats}},\ }\bibfield  {title} {\enquote
  {\bibinfo {title} {Relativistic field-theory spin and momentum in water
  waves},}\ }\href@noop {} {\bibfield  {journal} {\bibinfo  {journal}
  {arXiv:2009.03245}\ } (\bibinfo {year} {2020})}\BibitemShut {NoStop}%
\bibitem [{\citenamefont {Sugic}\ \emph {et~al.}(2020)\citenamefont {Sugic},
  \citenamefont {Dennis}, \citenamefont {Nori},\ and\ \citenamefont
  {Bliokh}}]{Sugic2020}%
  \BibitemOpen
  \bibfield  {author} {\bibinfo {author} {\bibfnamefont {D.}~\bibnamefont
  {Sugic}}, \bibinfo {author} {\bibfnamefont {M.~R.}\ \bibnamefont {Dennis}},
  \bibinfo {author} {\bibfnamefont {F.}~\bibnamefont {Nori}}, \ and\ \bibinfo
  {author} {\bibfnamefont {K.~Y.}\ \bibnamefont {Bliokh}},\ }\bibfield  {title}
  {\enquote {\bibinfo {title} {{Knotted polarizations and spin in 3D
  polychromatic waves}},}\ }\href@noop {} {\bibfield  {journal} {\bibinfo
  {journal} {Phys. Rev. Research}\ }\textbf {\bibinfo {volume} {2}},\ \bibinfo
  {pages} {042045(R)} (\bibinfo {year} {2020})}\BibitemShut {NoStop}%
\bibitem [{\citenamefont {Jones}(1973)}]{Jones1973}%
  \BibitemOpen
  \bibfield  {author} {\bibinfo {author} {\bibfnamefont {W.~L.}\ \bibnamefont
  {Jones}},\ }\bibfield  {title} {\enquote {\bibinfo {title} {Asymmetric
  wave-stress tensors and wave spin},}\ }\href@noop {} {\bibfield  {journal}
  {\bibinfo  {journal} {J. Fluid Mech.}\ }\textbf {\bibinfo {volume} {58}},\
  \bibinfo {pages} {737--747} (\bibinfo {year} {1973})}\BibitemShut {NoStop}%
\bibitem [{\citenamefont {Longuet-Higgins}(1980)}]{Longuet-Higgins1980}%
  \BibitemOpen
  \bibfield  {author} {\bibinfo {author} {\bibfnamefont {M.~S.}\ \bibnamefont
  {Longuet-Higgins}},\ }\bibfield  {title} {\enquote {\bibinfo {title} {Spin
  and angular momentum in gravity waves},}\ }\href@noop {} {\bibfield
  {journal} {\bibinfo  {journal} {J. Fluid Mech.}\ }\textbf {\bibinfo {volume}
  {97}},\ \bibinfo {pages} {1--25} (\bibinfo {year} {1980})}\BibitemShut
  {NoStop}%
\bibitem [{\citenamefont {Landau}\ and\ \citenamefont
  {Lifshitz}(1987)}]{LLfluid}%
  \BibitemOpen
  \bibfield  {author} {\bibinfo {author} {\bibfnamefont {L.~D.}\ \bibnamefont
  {Landau}}\ and\ \bibinfo {author} {\bibfnamefont {E.~M.}\ \bibnamefont
  {Lifshitz}},\ }\href@noop {} {\emph {\bibinfo {title} {Fluid Mechanics}}}\
  (\bibinfo  {publisher} {{Butterworth-Heinemann, Oxford}},\ \bibinfo {year}
  {1987})\BibitemShut {NoStop}%
\bibitem [{\citenamefont {Bliokh}\ and\ \citenamefont
  {Nori}(2015)}]{Bliokh2015PR}%
  \BibitemOpen
  \bibfield  {author} {\bibinfo {author} {\bibfnamefont {K.~Y.}\ \bibnamefont
  {Bliokh}}\ and\ \bibinfo {author} {\bibfnamefont {F.}~\bibnamefont {Nori}},\
  }\bibfield  {title} {\enquote {\bibinfo {title} {Transverse and longitudinal
  angular momenta of light},}\ }\href@noop {} {\bibfield  {journal} {\bibinfo
  {journal} {Phys. Rep.}\ }\textbf {\bibinfo {volume} {592}},\ \bibinfo {pages}
  {1--38} (\bibinfo {year} {2015})}\BibitemShut {NoStop}%
\bibitem [{\citenamefont {Aiello}\ \emph {et~al.}(2015)\citenamefont {Aiello},
  \citenamefont {Banzer}, \citenamefont {Neugebauer},\ and\ \citenamefont
  {Leuchs}}]{Aiello2015}%
  \BibitemOpen
  \bibfield  {author} {\bibinfo {author} {\bibfnamefont {A.}~\bibnamefont
  {Aiello}}, \bibinfo {author} {\bibfnamefont {P.}~\bibnamefont {Banzer}},
  \bibinfo {author} {\bibfnamefont {M.}~\bibnamefont {Neugebauer}}, \ and\
  \bibinfo {author} {\bibfnamefont {G.}~\bibnamefont {Leuchs}},\ }\bibfield
  {title} {\enquote {\bibinfo {title} {From transverse angular momentum to
  photonic wheels},}\ }\href@noop {} {\bibfield  {journal} {\bibinfo  {journal}
  {Nat. Photon.}\ }\textbf {\bibinfo {volume} {9}},\ \bibinfo {pages}
  {789--795} (\bibinfo {year} {2015})}\BibitemShut {NoStop}%
\bibitem [{\citenamefont
  {www.saddleback.edu/faculty/jrepka/notes/waves.html}()}]{surfacewave}%
  \BibitemOpen
  \bibfield  {author} {\bibinfo {author} {\bibnamefont
  {www.saddleback.edu/faculty/jrepka/notes/waves.html}},\ }\href@noop {}
  {}\BibitemShut {NoStop}%
\bibitem [{\citenamefont {Dennis}\ \emph {et~al.}(2009)\citenamefont {Dennis},
  \citenamefont {O'Holleran},\ and\ \citenamefont {Padgett}}]{Dennis2009}%
  \BibitemOpen
  \bibfield  {author} {\bibinfo {author} {\bibfnamefont {M.~R.}\ \bibnamefont
  {Dennis}}, \bibinfo {author} {\bibfnamefont {K.}~\bibnamefont {O'Holleran}},
  \ and\ \bibinfo {author} {\bibfnamefont {M.~J.}\ \bibnamefont {Padgett}},\
  }\bibfield  {title} {\enquote {\bibinfo {title} {Singular optics: Optical
  vortices and polarization singularities},}\ }\href@noop {} {\bibfield
  {journal} {\bibinfo  {journal} {Prog. Opt.}\ }\textbf {\bibinfo {volume}
  {53}},\ \bibinfo {pages} {293--363} (\bibinfo {year} {2009})}\BibitemShut
  {NoStop}%
\bibitem [{\citenamefont {Eggers}(2018)}]{eggers_PRFluids_2018}%
  \BibitemOpen
  \bibfield  {author} {\bibinfo {author} {\bibfnamefont {J.}~\bibnamefont
  {Eggers}},\ }\bibfield  {title} {\enquote {\bibinfo {title} {Role of
  singularities in hydrodynamics},}\ }\href@noop {} {\bibfield  {journal}
  {\bibinfo  {journal} {Phys. Rev. Fluids}\ }\textbf {\bibinfo {volume} {3}},\
  \bibinfo {pages} {110503} (\bibinfo {year} {2018})}\BibitemShut {NoStop}%
\bibitem [{\citenamefont {Moffatt}(2019)}]{moffatt_PRFluids_2019}%
  \BibitemOpen
  \bibfield  {author} {\bibinfo {author} {\bibfnamefont {H.~K.}\ \bibnamefont
  {Moffatt}},\ }\bibfield  {title} {\enquote {\bibinfo {title} {Singularities
  in fluid mechanics},}\ }\href@noop {} {\bibfield  {journal} {\bibinfo
  {journal} {Phys. Rev. Fluids}\ }\textbf {\bibinfo {volume} {4}},\ \bibinfo
  {pages} {110502} (\bibinfo {year} {2019})}\BibitemShut {NoStop}%
\bibitem [{\citenamefont {Kleckner}\ and\ \citenamefont
  {Irvine}(2013)}]{kleckner_natphys_2013}%
  \BibitemOpen
  \bibfield  {author} {\bibinfo {author} {\bibfnamefont {D.}~\bibnamefont
  {Kleckner}}\ and\ \bibinfo {author} {\bibfnamefont {W.~T.~M.}\ \bibnamefont
  {Irvine}},\ }\bibfield  {title} {\enquote {\bibinfo {title} {Creation and
  dynamics of knotted vortices},}\ }\href@noop {} {\bibfield  {journal}
  {\bibinfo  {journal} {Nat. Phys.}\ }\textbf {\bibinfo {volume} {9}},\
  \bibinfo {pages} {253--258} (\bibinfo {year} {2013})}\BibitemShut {NoStop}%
\bibitem [{\citenamefont {Zhang}\ \emph {et~al.}(2020)\citenamefont {Zhang},
  \citenamefont {Zhang}, \citenamefont {Liao}, \citenamefont {Zhou},
  \citenamefont {Li}, \citenamefont {Hu},\ and\ \citenamefont
  {Zhang}}]{zhang_natcommun_2020}%
  \BibitemOpen
  \bibfield  {author} {\bibinfo {author} {\bibfnamefont {H.}~\bibnamefont
  {Zhang}}, \bibinfo {author} {\bibfnamefont {W.}~\bibnamefont {Zhang}},
  \bibinfo {author} {\bibfnamefont {Y.}~\bibnamefont {Liao}}, \bibinfo {author}
  {\bibfnamefont {X.}~\bibnamefont {Zhou}}, \bibinfo {author} {\bibfnamefont
  {J.}~\bibnamefont {Li}}, \bibinfo {author} {\bibfnamefont {G.}~\bibnamefont
  {Hu}}, \ and\ \bibinfo {author} {\bibfnamefont {X.}~\bibnamefont {Zhang}},\
  }\bibfield  {title} {\enquote {\bibinfo {title} {Creation of acoustic vortex
  knots},}\ }\href@noop {} {\bibfield  {journal} {\bibinfo  {journal} {Nat.
  Commun.}\ }\textbf {\bibinfo {volume} {11}},\ \bibinfo {pages} {3956}
  (\bibinfo {year} {2020})}\BibitemShut {NoStop}%
\bibitem [{\citenamefont {Chen}\ and\ \citenamefont
  {Meiners}(2004)}]{chen_apl_2004}%
  \BibitemOpen
  \bibfield  {author} {\bibinfo {author} {\bibfnamefont {H.}~\bibnamefont
  {Chen}}\ and\ \bibinfo {author} {\bibfnamefont {J.-C.}\ \bibnamefont
  {Meiners}},\ }\bibfield  {title} {\enquote {\bibinfo {title} {Topologic
  mixing on a microfluidic chip},}\ }\href@noop {} {\bibfield  {journal}
  {\bibinfo  {journal} {Appl. Phys. Lett.}\ }\textbf {\bibinfo {volume} {84}},\
  \bibinfo {pages} {2193--2195} (\bibinfo {year} {2004})}\BibitemShut {NoStop}%
\bibitem [{\citenamefont {Goldstein}\ \emph {et~al.}(2010)\citenamefont
  {Goldstein}, \citenamefont {Moffatt}, \citenamefont {Pesci},\ and\
  \citenamefont {Ricca}}]{goldstein_pnas_2010}%
  \BibitemOpen
  \bibfield  {author} {\bibinfo {author} {\bibfnamefont {R.~E.}\ \bibnamefont
  {Goldstein}}, \bibinfo {author} {\bibfnamefont {H.~K.}\ \bibnamefont
  {Moffatt}}, \bibinfo {author} {\bibfnamefont {A.~I.}\ \bibnamefont {Pesci}},
  \ and\ \bibinfo {author} {\bibfnamefont {R.~L.}\ \bibnamefont {Ricca}},\
  }\bibfield  {title} {\enquote {\bibinfo {title} {{Soap-film M{\"o}bius strip
  changes topology with a twist singularity}},}\ }\href@noop {} {\bibfield
  {journal} {\bibinfo  {journal} {Proc. Natl. Acad. Sci. USA}\ }\textbf
  {\bibinfo {volume} {107}},\ \bibinfo {pages} {21979--21984} (\bibinfo {year}
  {2010})}\BibitemShut {NoStop}%
\bibitem [{\citenamefont {Whewell}(1836)}]{Whewell}%
  \BibitemOpen
  \bibfield  {author} {\bibinfo {author} {\bibfnamefont {W.}~\bibnamefont
  {Whewell}},\ }\bibfield  {title} {\enquote {\bibinfo {title} {On the results
  of an extensive series of tide observations},}\ }\href@noop {} {\bibfield
  {journal} {\bibinfo  {journal} {Phil. Trans. Roy. Soc. Lond.}\ ,\ \bibinfo
  {pages} {289--307}} (\bibinfo {year} {1836})}\BibitemShut {NoStop}%
\bibitem [{\citenamefont {Hansen}(1952)}]{Hansen}%
  \BibitemOpen
  \bibfield  {author} {\bibinfo {author} {\bibfnamefont {W.}~\bibnamefont
  {Hansen}},\ }\href@noop {} {\emph {\bibinfo {title} {{Gezeiten und
  Gezeitenstr{\"o}me der halbt{\"a}gigen Hauptmondtide M2 in der Nordsee}}}}\
  (\bibinfo  {publisher} {Hamburg: Deutsche Hydrographisches Institut},\
  \bibinfo {year} {1952})\BibitemShut {NoStop}%
\bibitem [{\citenamefont {Berry}(2001)}]{Berry2001}%
  \BibitemOpen
  \bibfield  {author} {\bibinfo {author} {\bibfnamefont {M.~V.}\ \bibnamefont
  {Berry}},\ }\bibfield  {title} {\enquote {\bibinfo {title} {Geometry of phase
  and polarization singularities, illustrated by edge diffraction and the
  tides},}\ }\href@noop {} {\bibfield  {journal} {\bibinfo  {journal} {{Proc.
  SPIE}}\ }\textbf {\bibinfo {volume} {4403}},\ \bibinfo {pages} {1} (\bibinfo
  {year} {2001})}\BibitemShut {NoStop}%
\bibitem [{\citenamefont {Nye}\ \emph {et~al.}(1988)\citenamefont {Nye},
  \citenamefont {Hajnal},\ and\ \citenamefont {Hannay}}]{Nye1988}%
  \BibitemOpen
  \bibfield  {author} {\bibinfo {author} {\bibfnamefont {J.~F.}\ \bibnamefont
  {Nye}}, \bibinfo {author} {\bibfnamefont {J.~V.}\ \bibnamefont {Hajnal}}, \
  and\ \bibinfo {author} {\bibfnamefont {J.~H.}\ \bibnamefont {Hannay}},\
  }\bibfield  {title} {\enquote {\bibinfo {title} {Phase saddles and
  dislocations in two-dimensional waves such as the tides},}\ }\href@noop {}
  {\bibfield  {journal} {\bibinfo  {journal} {Proc. Roy. Soc. Lond. A A}\
  }\textbf {\bibinfo {volume} {417}},\ \bibinfo {pages} {7--20} (\bibinfo
  {year} {1988})}\BibitemShut {NoStop}%
\bibitem [{\citenamefont {Nye}\ and\ \citenamefont {Hajnal}(1987)}]{Nye1987}%
  \BibitemOpen
  \bibfield  {author} {\bibinfo {author} {\bibfnamefont {J.~F.}\ \bibnamefont
  {Nye}}\ and\ \bibinfo {author} {\bibfnamefont {J.~V.}\ \bibnamefont
  {Hajnal}},\ }\bibfield  {title} {\enquote {\bibinfo {title} {The wave
  structure of monochromatic electromagnetic radiation},}\ }\href@noop {}
  {\bibfield  {journal} {\bibinfo  {journal} {Proc. Roy. Soc. Lond. A}\
  }\textbf {\bibinfo {volume} {409}},\ \bibinfo {pages} {21--36} (\bibinfo
  {year} {1987})}\BibitemShut {NoStop}%
\bibitem [{\citenamefont {Soskin}\ and\ \citenamefont
  {Vasnetsov}(2001)}]{Soskin2001}%
  \BibitemOpen
  \bibfield  {author} {\bibinfo {author} {\bibfnamefont {M.~S.}\ \bibnamefont
  {Soskin}}\ and\ \bibinfo {author} {\bibfnamefont {M.~V.}\ \bibnamefont
  {Vasnetsov}},\ }\bibfield  {title} {\enquote {\bibinfo {title} {Singular
  optics},}\ }\href@noop {} {\bibfield  {journal} {\bibinfo  {journal} {Prog.
  Opt.}\ }\textbf {\bibinfo {volume} {42}},\ \bibinfo {pages} {219--276}
  (\bibinfo {year} {2001})}\BibitemShut {NoStop}%
\bibitem [{\citenamefont {Berry}\ and\ \citenamefont
  {Dennis}(2001)}]{BerryDennis2001}%
  \BibitemOpen
  \bibfield  {author} {\bibinfo {author} {\bibfnamefont {M.~V.}\ \bibnamefont
  {Berry}}\ and\ \bibinfo {author} {\bibfnamefont {M.~R.}\ \bibnamefont
  {Dennis}},\ }\bibfield  {title} {\enquote {\bibinfo {title} {Polarization
  singularities in isotropic random vector waves},}\ }\href@noop {} {\bibfield
  {journal} {\bibinfo  {journal} {Proc. R. Soc. Lond. A}\ }\textbf {\bibinfo
  {volume} {457}},\ \bibinfo {pages} {141} (\bibinfo {year}
  {2001})}\BibitemShut {NoStop}%
\bibitem [{\citenamefont {Nye}(1983)}]{Nye1983}%
  \BibitemOpen
  \bibfield  {author} {\bibinfo {author} {\bibfnamefont {J.~F.}\ \bibnamefont
  {Nye}},\ }\bibfield  {title} {\enquote {\bibinfo {title} {Lines of circular
  polarization in electromagnetic wave fields},}\ }\href@noop {} {\bibfield
  {journal} {\bibinfo  {journal} {Proc. Roy. Soc. Lond. A}\ }\textbf {\bibinfo
  {volume} {389}},\ \bibinfo {pages} {279--290} (\bibinfo {year}
  {1983})}\BibitemShut {NoStop}%
\bibitem [{\citenamefont {Hajnal}(1987)}]{Hajnal1987}%
  \BibitemOpen
  \bibfield  {author} {\bibinfo {author} {\bibfnamefont {J.~V.}\ \bibnamefont
  {Hajnal}},\ }\bibfield  {title} {\enquote {\bibinfo {title} {{Singularities
  in the transverse fields of electromagnetic waves. II. Observations on the
  electric field}},}\ }\href@noop {} {\bibfield  {journal} {\bibinfo  {journal}
  {Proc. R. Soc. Lond. A}\ }\textbf {\bibinfo {volume} {414}},\ \bibinfo
  {pages} {447--468} (\bibinfo {year} {1987})}\BibitemShut {NoStop}%
\bibitem [{\citenamefont {Bliokh}\ \emph {et~al.}(2019)\citenamefont {Bliokh},
  \citenamefont {Alonso},\ and\ \citenamefont {Dennis}}]{BAD2019}%
  \BibitemOpen
  \bibfield  {author} {\bibinfo {author} {\bibfnamefont {K.~Y.}\ \bibnamefont
  {Bliokh}}, \bibinfo {author} {\bibfnamefont {M.~A.}\ \bibnamefont {Alonso}},
  \ and\ \bibinfo {author} {\bibfnamefont {M.~R.}\ \bibnamefont {Dennis}},\
  }\bibfield  {title} {\enquote {\bibinfo {title} {{Geometric phases in 2D and
  3D polarized fields: geometrical, dynamical, and topological aspects}},}\
  }\href@noop {} {\bibfield  {journal} {\bibinfo  {journal} {Rep. Prog. Phys.}\
  }\textbf {\bibinfo {volume} {82}},\ \bibinfo {pages} {122401} (\bibinfo
  {year} {2019})}\BibitemShut {NoStop}%
\bibitem [{\citenamefont {Freund}(2010{\natexlab{a}})}]{Freund2010}%
  \BibitemOpen
  \bibfield  {author} {\bibinfo {author} {\bibfnamefont {I.}~\bibnamefont
  {Freund}},\ }\bibfield  {title} {\enquote {\bibinfo {title} {{Optical
  M{\"o}bius strips in three-dimensional ellipse fields: I. Lines of circular
  polarization}},}\ }\href@noop {} {\bibfield  {journal} {\bibinfo  {journal}
  {Opt. Commun.}\ }\textbf {\bibinfo {volume} {283}},\ \bibinfo {pages} {1--15}
  (\bibinfo {year} {2010}{\natexlab{a}})}\BibitemShut {NoStop}%
\bibitem [{\citenamefont {Freund}(2010{\natexlab{b}})}]{Freund2010II}%
  \BibitemOpen
  \bibfield  {author} {\bibinfo {author} {\bibfnamefont {I.}~\bibnamefont
  {Freund}},\ }\bibfield  {title} {\enquote {\bibinfo {title} {{Multitwist
  optical M{\"o}bius strips}},}\ }\href@noop {} {\bibfield  {journal} {\bibinfo
   {journal} {Opt. Lett.}\ }\textbf {\bibinfo {volume} {35}},\ \bibinfo {pages}
  {148--150} (\bibinfo {year} {2010}{\natexlab{b}})}\BibitemShut {NoStop}%
\bibitem [{\citenamefont {Dennis}(2011)}]{Dennis2011}%
  \BibitemOpen
  \bibfield  {author} {\bibinfo {author} {\bibfnamefont {M.~R.}\ \bibnamefont
  {Dennis}},\ }\bibfield  {title} {\enquote {\bibinfo {title} {Fermionic
  out-of-plane structure of polarization singularities},}\ }\href@noop {}
  {\bibfield  {journal} {\bibinfo  {journal} {Opt. Lett.}\ }\textbf {\bibinfo
  {volume} {36}},\ \bibinfo {pages} {3765--3767} (\bibinfo {year}
  {2011})}\BibitemShut {NoStop}%
\bibitem [{\citenamefont {Bauer}\ \emph {et~al.}(2015)\citenamefont {Bauer},
  \citenamefont {Banzer}, \citenamefont {Karimi}, \citenamefont {Orlov},
  \citenamefont {Rubano}, \citenamefont {Marrucci}, \citenamefont {Santamato},
  \citenamefont {Boyd},\ and\ \citenamefont {Leuchs}}]{Bauer2015}%
  \BibitemOpen
  \bibfield  {author} {\bibinfo {author} {\bibfnamefont {T.}~\bibnamefont
  {Bauer}}, \bibinfo {author} {\bibfnamefont {P.}~\bibnamefont {Banzer}},
  \bibinfo {author} {\bibfnamefont {E.}~\bibnamefont {Karimi}}, \bibinfo
  {author} {\bibfnamefont {S.}~\bibnamefont {Orlov}}, \bibinfo {author}
  {\bibfnamefont {A.}~\bibnamefont {Rubano}}, \bibinfo {author} {\bibfnamefont
  {L.}~\bibnamefont {Marrucci}}, \bibinfo {author} {\bibfnamefont
  {E.}~\bibnamefont {Santamato}}, \bibinfo {author} {\bibfnamefont {R.~W.}\
  \bibnamefont {Boyd}}, \ and\ \bibinfo {author} {\bibfnamefont
  {G.}~\bibnamefont {Leuchs}},\ }\bibfield  {title} {\enquote {\bibinfo {title}
  {{Observation of optical polarization M{\"o}bius strips}},}\ }\href@noop {}
  {\bibfield  {journal} {\bibinfo  {journal} {Science}\ }\textbf {\bibinfo
  {volume} {347}},\ \bibinfo {pages} {964--966} (\bibinfo {year}
  {2015})}\BibitemShut {NoStop}%
\bibitem [{\citenamefont {Galvez}\ \emph {et~al.}(2017)\citenamefont {Galvez},
  \citenamefont {Dutta}, \citenamefont {Beach}, \citenamefont {Zeosky},
  \citenamefont {Jones},\ and\ \citenamefont {Khajavi}}]{Galvez2017}%
  \BibitemOpen
  \bibfield  {author} {\bibinfo {author} {\bibfnamefont {E.~J.}\ \bibnamefont
  {Galvez}}, \bibinfo {author} {\bibfnamefont {I.}~\bibnamefont {Dutta}},
  \bibinfo {author} {\bibfnamefont {K.}~\bibnamefont {Beach}}, \bibinfo
  {author} {\bibfnamefont {J.~J.}\ \bibnamefont {Zeosky}}, \bibinfo {author}
  {\bibfnamefont {J.~A.}\ \bibnamefont {Jones}}, \ and\ \bibinfo {author}
  {\bibfnamefont {B.}~\bibnamefont {Khajavi}},\ }\bibfield  {title} {\enquote
  {\bibinfo {title} {{Multitwist M{\"o}bius strips and twisted ribbons in the
  polarization of paraxial light beams}},}\ }\href@noop {} {\bibfield
  {journal} {\bibinfo  {journal} {Sci. Rep.}\ }\textbf {\bibinfo {volume}
  {7}},\ \bibinfo {pages} {13653} (\bibinfo {year} {2017})}\BibitemShut
  {NoStop}%
\bibitem [{\citenamefont {Garcia-Etxarri}(2017)}]{Garcia2017}%
  \BibitemOpen
  \bibfield  {author} {\bibinfo {author} {\bibfnamefont {A.}~\bibnamefont
  {Garcia-Etxarri}},\ }\bibfield  {title} {\enquote {\bibinfo {title} {{Optical
  polarization M{\"o}bius strips on all-Dielectric optical scatterers}},}\
  }\href@noop {} {\bibfield  {journal} {\bibinfo  {journal} {{ACS Photon.}}\
  }\textbf {\bibinfo {volume} {4}},\ \bibinfo {pages} {1159--1164} (\bibinfo
  {year} {2017})}\BibitemShut {NoStop}%
\bibitem [{\citenamefont {Kreismann}\ and\ \citenamefont
  {Hentschel}(2018)}]{Kreismann2017}%
  \BibitemOpen
  \bibfield  {author} {\bibinfo {author} {\bibfnamefont {J.}~\bibnamefont
  {Kreismann}}\ and\ \bibinfo {author} {\bibfnamefont {M.}~\bibnamefont
  {Hentschel}},\ }\bibfield  {title} {\enquote {\bibinfo {title} {{The optical
  M\"obius strip cavity: Tailoring geometric phases and far fields}},}\
  }\href@noop {} {\bibfield  {journal} {\bibinfo  {journal} {EPL}\ }\textbf
  {\bibinfo {volume} {121}},\ \bibinfo {pages} {24001} (\bibinfo {year}
  {2018})}\BibitemShut {NoStop}%
\bibitem [{\citenamefont {Bauer}\ \emph {et~al.}(2019)\citenamefont {Bauer},
  \citenamefont {Banzer}, \citenamefont {Bouchard}, \citenamefont {Orlov},
  \citenamefont {Marrucci}, \citenamefont {amd R.~W.~Boyd}, \citenamefont
  {Karimi},\ and\ \citenamefont {Leuchs}}]{Bauer2019}%
  \BibitemOpen
  \bibfield  {author} {\bibinfo {author} {\bibfnamefont {T.}~\bibnamefont
  {Bauer}}, \bibinfo {author} {\bibfnamefont {P.}~\bibnamefont {Banzer}},
  \bibinfo {author} {\bibfnamefont {F.}~\bibnamefont {Bouchard}}, \bibinfo
  {author} {\bibfnamefont {S.}~\bibnamefont {Orlov}}, \bibinfo {author}
  {\bibfnamefont {L.}~\bibnamefont {Marrucci}}, \bibinfo {author}
  {\bibfnamefont {E.~Santamato}\ \bibnamefont {amd R.~W.~Boyd}}, \bibinfo
  {author} {\bibfnamefont {E.}~\bibnamefont {Karimi}}, \ and\ \bibinfo {author}
  {\bibfnamefont {G.}~\bibnamefont {Leuchs}},\ }\bibfield  {title} {\enquote
  {\bibinfo {title} {Multi-twist polarization ribbon topologies in
  highly-confined optical fields},}\ }\href@noop {} {\bibfield  {journal}
  {\bibinfo  {journal} {New J. Phys.}\ }\textbf {\bibinfo {volume} {21}},\
  \bibinfo {pages} {053020} (\bibinfo {year} {2019})}\BibitemShut {NoStop}%
\bibitem [{\citenamefont {Tekce}\ \emph {et~al.}(2019)\citenamefont {Tekce},
  \citenamefont {Otte},\ and\ \citenamefont {Denz}}]{Tekce2019}%
  \BibitemOpen
  \bibfield  {author} {\bibinfo {author} {\bibfnamefont {K.}~\bibnamefont
  {Tekce}}, \bibinfo {author} {\bibfnamefont {E.}~\bibnamefont {Otte}}, \ and\
  \bibinfo {author} {\bibfnamefont {C.}~\bibnamefont {Denz}},\ }\bibfield
  {title} {\enquote {\bibinfo {title} {{Optical singularities and M{\"o}bius
  strip arrays in tailored non-paraxial light fields}},}\ }\href@noop {}
  {\bibfield  {journal} {\bibinfo  {journal} {Opt. Express}\ }\textbf {\bibinfo
  {volume} {27}},\ \bibinfo {pages} {29685--29696} (\bibinfo {year}
  {2019})}\BibitemShut {NoStop}%
\bibitem [{\citenamefont {Taylor}(1976)}]{Taylor1976}%
  \BibitemOpen
  \bibfield  {author} {\bibinfo {author} {\bibfnamefont {K.~J.}\ \bibnamefont
  {Taylor}},\ }\bibfield  {title} {\enquote {\bibinfo {title} {Absolute
  measurement of acoustic particle velocity},}\ }\href@noop {} {\bibfield
  {journal} {\bibinfo  {journal} {J. Acoust. Soc. Am.}\ }\textbf {\bibinfo
  {volume} {59}},\ \bibinfo {pages} {691--694} (\bibinfo {year}
  {1976})}\BibitemShut {NoStop}%
\bibitem [{\citenamefont {Gabrielson}\ \emph {et~al.}(1995)\citenamefont
  {Gabrielson}, \citenamefont {Gardner},\ and\ \citenamefont
  {Garrett}}]{Gabrielson1995}%
  \BibitemOpen
  \bibfield  {author} {\bibinfo {author} {\bibfnamefont {T.~B.}\ \bibnamefont
  {Gabrielson}}, \bibinfo {author} {\bibfnamefont {D.~L.}\ \bibnamefont
  {Gardner}}, \ and\ \bibinfo {author} {\bibfnamefont {S.~L.}\ \bibnamefont
  {Garrett}},\ }\bibfield  {title} {\enquote {\bibinfo {title} {A simple
  neutrally buoyant sensor for direct measurement of particle velocity and
  intensity in water},}\ }\href@noop {} {\bibfield  {journal} {\bibinfo
  {journal} {J. Acoust. Soc. Am.}\ }\textbf {\bibinfo {volume} {97}},\ \bibinfo
  {pages} {2227--2237} (\bibinfo {year} {1995})}\BibitemShut {NoStop}%
\bibitem [{\citenamefont {Francois}\ \emph {et~al.}(2017)\citenamefont
  {Francois}, \citenamefont {Xia}, \citenamefont {Punzmann}, \citenamefont
  {Fontana},\ and\ \citenamefont {Shats}}]{Francois2017}%
  \BibitemOpen
  \bibfield  {author} {\bibinfo {author} {\bibfnamefont {N.}~\bibnamefont
  {Francois}}, \bibinfo {author} {\bibfnamefont {H.}~\bibnamefont {Xia}},
  \bibinfo {author} {\bibfnamefont {H.}~\bibnamefont {Punzmann}}, \bibinfo
  {author} {\bibfnamefont {P.~W.}\ \bibnamefont {Fontana}}, \ and\ \bibinfo
  {author} {\bibfnamefont {M.}~\bibnamefont {Shats}},\ }\bibfield  {title}
  {\enquote {\bibinfo {title} {Wave-based liquid-interface metamaterials},}\
  }\href@noop {} {\bibfield  {journal} {\bibinfo  {journal} {Nat. Commun.}\
  }\textbf {\bibinfo {volume} {8}},\ \bibinfo {pages} {14325} (\bibinfo {year}
  {2017})}\BibitemShut {NoStop}%
\bibitem [{\citenamefont {Wei}\ and\ \citenamefont
  {Rodriguez-Fortuno}(2020)}]{Wei2020}%
  \BibitemOpen
  \bibfield  {author} {\bibinfo {author} {\bibfnamefont {L.}~\bibnamefont
  {Wei}}\ and\ \bibinfo {author} {\bibfnamefont {F.~J.}\ \bibnamefont
  {Rodriguez-Fortuno}},\ }\bibfield  {title} {\enquote {\bibinfo {title}
  {Far-field and near-field directionality in acoustic scattering},}\
  }\href@noop {} {\bibfield  {journal} {\bibinfo  {journal} {New J. Phys.}\
  }\textbf {\bibinfo {volume} {22}},\ \bibinfo {pages} {083016} (\bibinfo
  {year} {2020})}\BibitemShut {NoStop}%
\bibitem [{\citenamefont {Long}\ \emph
  {et~al.}(2020{\natexlab{b}})\citenamefont {Long}, \citenamefont {Ge},
  \citenamefont {Zhang}, \citenamefont {Xu}, \citenamefont {Ren}, \citenamefont
  {Lu}, \citenamefont {Bao}, \citenamefont {Chen},\ and\ \citenamefont
  {Chen}}]{Long2020_II}%
  \BibitemOpen
  \bibfield  {author} {\bibinfo {author} {\bibfnamefont {Y.}~\bibnamefont
  {Long}}, \bibinfo {author} {\bibfnamefont {H.}~\bibnamefont {Ge}}, \bibinfo
  {author} {\bibfnamefont {D.}~\bibnamefont {Zhang}}, \bibinfo {author}
  {\bibfnamefont {X.}~\bibnamefont {Xu}}, \bibinfo {author} {\bibfnamefont
  {J.}~\bibnamefont {Ren}}, \bibinfo {author} {\bibfnamefont {M.-H.}\
  \bibnamefont {Lu}}, \bibinfo {author} {\bibfnamefont {M.}~\bibnamefont
  {Bao}}, \bibinfo {author} {\bibfnamefont {H.}~\bibnamefont {Chen}}, \ and\
  \bibinfo {author} {\bibfnamefont {Y.-F.}\ \bibnamefont {Chen}},\ }\bibfield
  {title} {\enquote {\bibinfo {title} {Symmetry selective directionality in
  near-field acoustics},}\ }\href@noop {} {\bibfield  {journal} {\bibinfo
  {journal} {Natl. Sci. Rev.}\ }\textbf {\bibinfo {volume} {7}},\ \bibinfo
  {pages} {1024--135} (\bibinfo {year} {2020}{\natexlab{b}})}\BibitemShut
  {NoStop}%
\bibitem [{\citenamefont {Dennis}(2002)}]{mrd2002}%
  \BibitemOpen
  \bibfield  {author} {\bibinfo {author} {\bibfnamefont {M.~R.}\ \bibnamefont
  {Dennis}},\ }\bibfield  {title} {\enquote {\bibinfo {title} {Polarization
  singularities in paraxial vector fields: morphology and statistics},}\
  }\href@noop {} {\bibfield  {journal} {\bibinfo  {journal} {Opt. Commun.}\
  }\textbf {\bibinfo {volume} {213}},\ \bibinfo {pages} {201--221} (\bibinfo
  {year} {2002})}\BibitemShut {NoStop}%
\bibitem [{\citenamefont {Freund}(2020{\natexlab{a}})}]{Freund2020}%
  \BibitemOpen
  \bibfield  {author} {\bibinfo {author} {\bibfnamefont {I.}~\bibnamefont
  {Freund}},\ }\bibfield  {title} {\enquote {\bibinfo {title} {{Polarization
  M{\"o}bius strips on elliptical paths in three-dimensional optical
  fields}},}\ }\href@noop {} {\bibfield  {journal} {\bibinfo  {journal} {Opt.
  Lett.}\ }\textbf {\bibinfo {volume} {45}},\ \bibinfo {pages} {3333--3336}
  (\bibinfo {year} {2020}{\natexlab{a}})}\BibitemShut {NoStop}%
\bibitem [{\citenamefont {Freund}(2020{\natexlab{b}})}]{Freund2020_II}%
  \BibitemOpen
  \bibfield  {author} {\bibinfo {author} {\bibfnamefont {I.}~\bibnamefont
  {Freund}},\ }\bibfield  {title} {\enquote {\bibinfo {title} {Twisted ribbon
  carousels in random, three-dimensional optical fields},}\ }\href@noop {}
  {\bibfield  {journal} {\bibinfo  {journal} {Opt. Lett.}\ }\textbf {\bibinfo
  {volume} {45}},\ \bibinfo {pages} {5905--5908} (\bibinfo {year}
  {2020}{\natexlab{b}})}\BibitemShut {NoStop}%
\bibitem [{\citenamefont {Nye}(1999)}]{Nye_book}%
  \BibitemOpen
  \bibfield  {author} {\bibinfo {author} {\bibfnamefont {J.}~\bibnamefont
  {Nye}},\ }\href@noop {} {\emph {\bibinfo {title} {Natural focusing and fine
  structure of light}}}\ (\bibinfo  {publisher} {{IOP Publishing, Bristol}},\
  \bibinfo {year} {1999})\BibitemShut {NoStop}%
\bibitem [{\citenamefont {Dennis}(2004)}]{Dennis2004_II}%
  \BibitemOpen
  \bibfield  {author} {\bibinfo {author} {\bibfnamefont {M.~R.}\ \bibnamefont
  {Dennis}},\ }\bibfield  {title} {\enquote {\bibinfo {title} {Local phase
  structure of wave dislocation lines: twist and twirl},}\ }\href@noop {}
  {\bibfield  {journal} {\bibinfo  {journal} {J. Opt. A: Pure Appl. Opt.}\
  }\textbf {\bibinfo {volume} {6}},\ \bibinfo {pages} {S202--S208} (\bibinfo
  {year} {2004})}\BibitemShut {NoStop}%
\bibitem [{\citenamefont {Ray}(2001)}]{Ray2001}%
  \BibitemOpen
  \bibfield  {author} {\bibinfo {author} {\bibfnamefont {R.~D.}\ \bibnamefont
  {Ray}},\ }\bibfield  {title} {\enquote {\bibinfo {title} {Inversion of
  oceanic tidal currents from measured elevations},}\ }\href@noop {} {\bibfield
   {journal} {\bibinfo  {journal} {J. Mar. Syst.}\ }\textbf {\bibinfo {volume}
  {28}},\ \bibinfo {pages} {1--18} (\bibinfo {year} {2001})}\BibitemShut
  {NoStop}%
\bibitem [{\citenamefont {Ray}\ and\ \citenamefont {Egbert}(2004)}]{Ray2004}%
  \BibitemOpen
  \bibfield  {author} {\bibinfo {author} {\bibfnamefont {R.~D.}\ \bibnamefont
  {Ray}}\ and\ \bibinfo {author} {\bibfnamefont {G.~D.}\ \bibnamefont
  {Egbert}},\ }\bibfield  {title} {\enquote {\bibinfo {title} {{The Global S1
  Tide}},}\ }\href@noop {} {\bibfield  {journal} {\bibinfo  {journal} {J. Phys.
  Oceanogr.}\ }\textbf {\bibinfo {volume} {34}},\ \bibinfo {pages} {1922--1935}
  (\bibinfo {year} {2004})}\BibitemShut {NoStop}%
\end{thebibliography}%

\end{document}